\documentclass[preprintnumbers,aps,prb,twocolumn]{revtex4}

\usepackage{graphicx}
\usepackage{dcolumn} 
\usepackage{amsmath}
\usepackage{exscale}
\usepackage{bm}      
\usepackage{color}
\usepackage{epsfig,psfrag,subfigure,subeqnarray}
\usepackage{bbm}
\usepackage{amsbsy}
\usepackage{amssymb}
\usepackage{enumitem}

\def\lan{\langle}
\def\ran{\rangle}
\def\va{\varepsilon}

\def\vk{{\bf k}}
\def\vK{{\bf K}}

\def\vr{{\bf r}}
\def\vu{{\bf u}}
\def\vQ{{\bf Q}}
\def\vq{{\bf q}}
\def\vp{{\bf p}}
\def\vR{{\bf R}}
\def\vP{{\bf P}}
\newcommand{\bd}{\begin{equation}}
\newcommand{\ed}{\end{equation}}
\newcommand{\be}{\begin{equation}}
\newcommand{\ee}{\end{equation}}
\newcommand{\bt}{\begin{split}}
\newcommand{\et}{\end{split}}

\newcommand{\bn}{\begin{align}}
\newcommand{\en}{\end{align}}

\newcommand{\bea}{\begin{eqnarray}}
\newcommand{\eea}{\end{eqnarray}}
\newcommand{\ba}{\begin{array}}
\newcommand{\ea}{\end{array}}
\newcommand{\nn}{\nonumber}

\begin{document}
\title {Electronic structure and absorption spectrum of biexciton obtained by using exciton basis}

\author{Shiue-Yuan Shiau$^{1,2}$, Monique Combescot$^3$  and Yia-Chung Chang$^1$}
\email{yiachang@gate.sinica.edu.tw}

\affiliation{$^1$Research Center for Applied Sciences, Academia Sinica, Taipei, 115 Taiwan}
\affiliation{$^2$ Department of Physics and National Center for Theoretical Sciences, National Cheng Kung University, Tainan, 701 Taiwan}
\affiliation{$^3$Institut des NanoSciences de Paris, Universit\'e Pierre et Marie Curie, CNRS, 4 place Jussieu, 75005 Paris}

\date{\today}

\begin{abstract}
We approach the biexciton Schr\"{o}dinger equation not through the free-carrier basis as usually done, but through the free-exciton basis, exciton-exciton interactions being treated according to the recently developed composite boson many-body formalism which allows an exact handling of carrier exchange between excitons, as induced by the Pauli exclusion principle. We numerically solve the resulting biexciton Schr\"{o}dinger equation with the exciton levels restricted to the ground state and we derive the biexciton ground state as well as the bound and unbound excited states as a function of hole-to-electron mass ratio. The biexciton ground-state energy we find, agrees reasonably well with variational results. Next, we use the obtained biexciton wave functions to calculate optical absorption in the presence of a dilute exciton gas in quantum well. We find a small asymmetric peak with a characteristic low-energy tail, identified with the biexciton ground state, and a set of large peaks associated with biexciton unbound states, i.e., exciton-exciton scattering states.
\end{abstract}


\maketitle

\section{Introduction\label{sec:biX1}}
The existence of composite particles in semiconductors has been predicted long ago\cite{lampert1958}. Bound states made of conduction band electrons and valence band holes result from the Coulomb attraction between these carriers. To name the simplest ones, these bound states are excitons ($X$) made of one electron plus one hole, trions ($X^{\mp}$) made of two electrons plus one hole, or two holes plus one electron, and biexcitons ($XX$) made of two electrons plus two holes. More exotic composite objects made of a large number of correlated fermion pairs, called ``electron-hole droplets", also exist, with a carrier density far larger than the one in which excitons are formed.\

The simplest composite particle, the exciton, made of one conduction electron and one valence hole, is very similar to an Hydrogen atom, if we neglect interband Coulomb processes\cite{MoniqueSSC2009}. Exciton has bound and unbound (scattering) states which can be analytically determined\cite{Elliott1957}. Exciton bound states appear as large narrow peaks in the photon absorption spectrum, the peak intensity depending on the so-called ``exciton oscillator strength". The reason bound exciton peaks are easy to observe is twofold: first, when a plane-wave photon with momentum $\vQ$ transforms into a bound exciton, the coupling is quite good because the center-of-mass motion of the bound exciton also is a plane wave with same momentum $\vQ$. Second, excitons, made of an even number of fermions, have a bosonic nature; so, they can be piled up all at the same energy, the absorption peak intensity increasing linearly with the density of excitons already present in the sample. \

Observation of composite particles like trions is more complex. It has been hampered for quite a long time, partly due to the small binding energies that trions have in bulk samples, at best one order of magnitude smaller than the exciton binding energy. Such binding energies are smaller than usual exciton line width; so, trion peaks fall on the side of exciton lines. This small trion binding energy can be physically understood by seeing the trion as an electron or a hole bound to an exciton. The effective attraction then is dipole-like which makes  interaction between exciton and free carrier much weaker than between elementary charges. A clear signature of trions has been obtained recently only in semiconductor quantum wells\cite{Kheng1993,Finkelstein1995,Shields1995,Buhmann1995}, the reduction of dimensionality increasing all binding energies as seen from the exciton energy which goes from $R_X^{\rm (3D)}$ in 3D to $R_X^{\rm (2D)}=4R_X^{\rm (3D)}$ in 2D, and to infinity in 1D. \

What makes bound trions hard to observe has also to be traced back to their oscillator strength which is one trion volume divided by one sample volume smaller than the exciton oscillator strength\cite{moniqSSC2003_2}. This drastic reduction factor can be physically understood as the probability for a photocreated exciton to localize over a trion volume, a free carrier initially spread over the whole sample. As a result, the trion peak commonly observed in heavily doped samples in which a large electron density exists, should not be interpreted as a signature of elementary trion, but rather as an exciton interacting in a coherent way with all the electrons present in the sample. Such a many-body effect is singular and leads to a broad absorption line, as experimentally shown\cite{moniqEP2005}. \

Mathematically, the derivation of trion eigenstates amounts to solving a three-body problem which has no known analytical solution. Attempts to tackle such a problem inevitably rely on some truncation scheme in addition to heavy numerics in order to possibly obtain satisfactory results. Recently, we showed how, using a physically relevant viewpoint, we can approach one trion as an exciton interacting with an electron\cite{Shiau2011}. We have constructed a Schr\"{o}dinger equation for trion using the electron-exciton basis and solved it by restricting this basis to the first few low-lying exciton levels,---which is reasonable since the energy scale for excitation of the exciton internal motion, of the order of one Rydberg, is much larger than the other energy scales. The resulting trion binding energies we find in 2D and 3D agree reasonably well with the most accurate variational results. One important advantage of this approach is to allow reaching the trion ground and excited states on equal footing, these excited states being out of reach from standard variational methods.  \

Following Lampert's prediction\cite{lampert1958}, an even more complex composite particle, the biexciton, has been observed in bulk materials such as CuCl [Ref.~\onlinecite{Nikitine1968}], Cu$_2$O [Ref.~\onlinecite{Lvov1971}], and AgBr [Ref.~\onlinecite{Pelant1976,Hulin1977}]. More recently, the biexciton binding energy has been measured in GaAs quantum well \cite{Miller1982} and found to be one order of magnitude larger than in bulk samples, a result supported by calculations done one year later\cite{Kleinman1983}. Since then, other aspects of biexcitons in confined structures, such as optical enhancement in biexciton formation\cite{Cingolani1988, Lovering1992} and the influence of dimensionality on the biexciton binding\cite{Miller1982,Birkedal1996,Euteneuer1997}, have been studied. Biexcitons in quantum wires have also been reported\cite{Crottini2002}.

In view of the successful application of the composite boson many-body formalism to trion\cite{Shiau2011}, we, in this work, go on along the same line to tackle biexciton. The biexciton problem {\it a priori} is an even more complex four-body problem, with two electrons $(e_1,e_2)$ and two holes $(h_1,h_2)$ involved. The idea is to start with two electron-hole pairs bound into two excitons by the strong electron-hole Coulomb attraction. The exciton-exciton attraction, although quite weak since it essentially is dipole-like, allows two free excitons to form a molecule with a binding energy substantially smaller than the exciton binding energy. To approach a system made of two electrons plus two holes, the exciton basis is physically quite appealing because the strong exciton binding energy is then included into the problem at the zeroth order. We are left with solving a Schr\"{o}dinger equation for the weaker biexciton binding energy. In this approach, the four-body system is pictured as two interacting excitons: one exciton is made of the $(e_1,h_1)$ pair, while the other is made of $(e_2,h_2)$ pair, these two excitons however exchanging their carriers to be possibly made of $(e_1,h_2)$ and $(e_2,h_1)$. Such a two-exciton picture could be thought, at first sight, to lead to an easy problem because of the weak exciton-exciton attraction compared with the strong electron-hole attraction. However, this weak attraction is the one responsible for two excitons to be bound into a molecule. So, in order to reach bound states and find the associated poles, this exciton-exciton interaction has to be treated in an exact way. \

With this goal in mind, we here construct a biexciton Schr\"{o}dinger equation in terms of the exciton basis using the recently developed composite boson many-body theory\cite{moniqPhysRep}. In much the same spirit as Feynman diagrams for elementary particles, this theory takes advantage of ``shiva diagrams" to visually identify many-body effects involved among composite particles. It moreover enables treating exactly carrier exchange which results from the indistinguisability of the fermionic components of these composite particles. By restricting the exciton levels to the ground state only, it becomes possible to numerically solve the biexciton Schr\"{o}dinger equation quite easily. The values we obtain for the biexciton ground state energies in 2D and 3D are in good agreement with variational results. One important advantage of the procedure is that the biexciton Schr\"{o}dinger equation can be cast into a generalized eigenvalue problem; so, we can reach bound and unbound excited states at once, with a single matrix diagonalization.  \

In a second step, we use the obtained biexciton relative motion wave functions to calculate the photon absorption spectrum in quantum wells, assumed to be exact 2D systems. Instead of considering biexciton as generated through two-photon absorption\cite{Hamamura1973,Arya1977,Ivanov1993,Hassan1993}, we here study one photocreated exciton interacting with a dilute exciton gas. Using similar arguments as those we used for bound trion, we find that the biexciton oscillator strength
is one biexciton volume divided by one sample volume smaller than the exciton oscillator strength. This would make observing the biexciton line very difficult. However, biexcitons, like excitons, are boson-like particles: They can thus be packed up all at the same energy level. As a result, the biexciton absorption line increases linearly with exciton density provided that the density is low enough to possibly neglect many-body effects between the photocreated exciton and the free excitons present in the sample. The calculated photon absorption spectrum shows a small peak, with a characteristic low-energy tail, originating from the biexciton molecular state, and large peaks centered on the exciton ground levels, which are associated with exciton-exciton scattering states. Both, the bound and unbound biexciton peak intensities decrease when the temperature increases. It can also be shown that the intensity of the absorption line for one biexciton made from a photocreated exciton and an exciton of the exciton gas, increases linearly with photon number and exciton density. This is in contrast to the biexciton absorption line associated with two-photon absorption which increases quadratically with photon number and thus becomes dominant at high laser intensity. \

The present paper is organized as follows:

In Sec.~\ref{sec:biX2}, we briefly discuss the relation which exists between biexciton written in the free-carrier basis, and biexciton written in the exciton basis. We also introduce the four commutators necessary to properly handle many-body effects involving composite excitons.\

In Sec.~\ref{sec:biX3}, we study triplet biexciton states made of same-spin electrons and same-spin holes and we derive the corresponding Schr\"{o}dinger equation.\

In Sec.~\ref{sec:biX4}, we study singlet and triplet biexciton states made of opposite-spin electrons and opposite-spin holes. These biexciton states are first constructed in terms of two free electrons plus two free holes, and then in terms of two free excitons, in order to reveal important parity relations. We then concentrate on singlet biexciton states with center-of-mass momentum equal to zero and we restrict the exciton levels to the ground state. This nicely reduces the biexciton Schr\"{o}dinger equation to a 1D integral equation.

In Sec.~\ref{sec:biX5}, we numerically solve this 1D integral equation to obtain the biexciton binding energies for the ground and excited states as a function of hole-to-electron mass ratio. We also show the biexciton relative motion wave functions for the bound state as well as for a few unbound states. Finally, we use these wave functions to calculate the photon absorption spectrum in the presence of a dilute exciton gas for various low temperatures.\

In the last section, we conclude.\

\begin{figure}[t]
\centering

\subfigure[]{
   \includegraphics[trim=4cm 6cm 4cm 4cm,clip,width=3.2in] {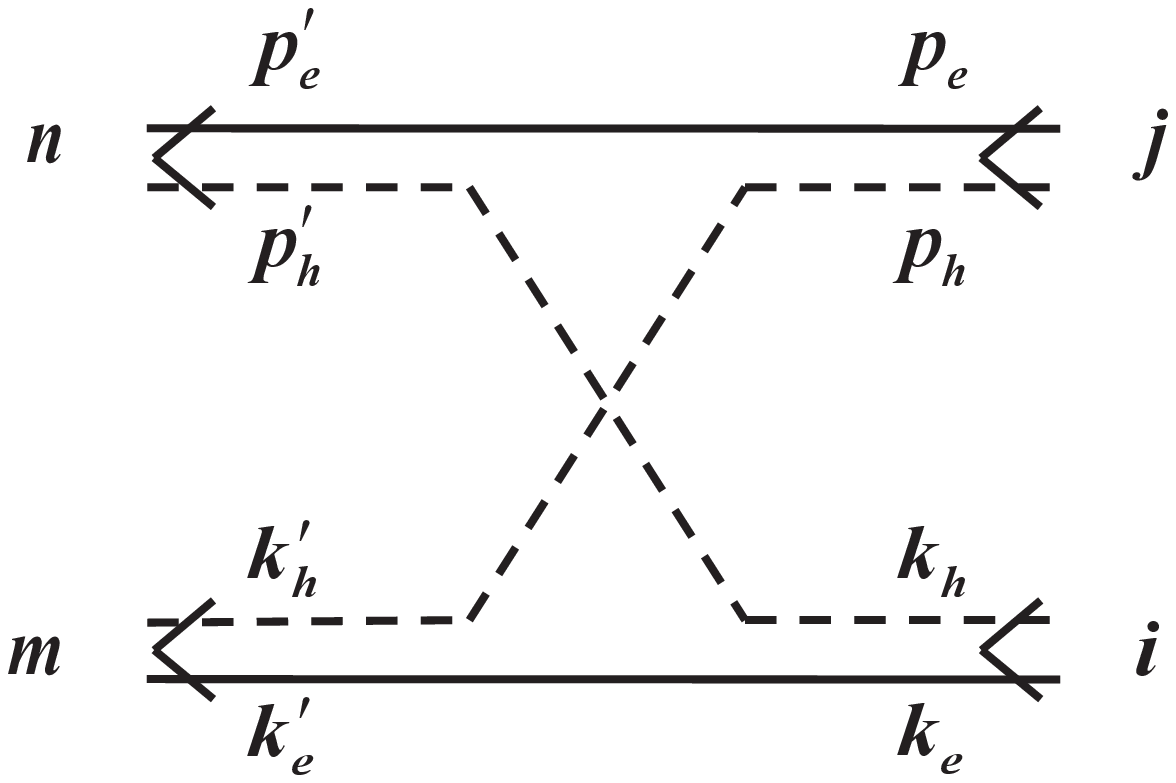}
   \label{fig:fig1a}
 }

 \subfigure[]{
   \includegraphics[trim=4cm 6cm 4cm 4cm,clip,width=3.2in] {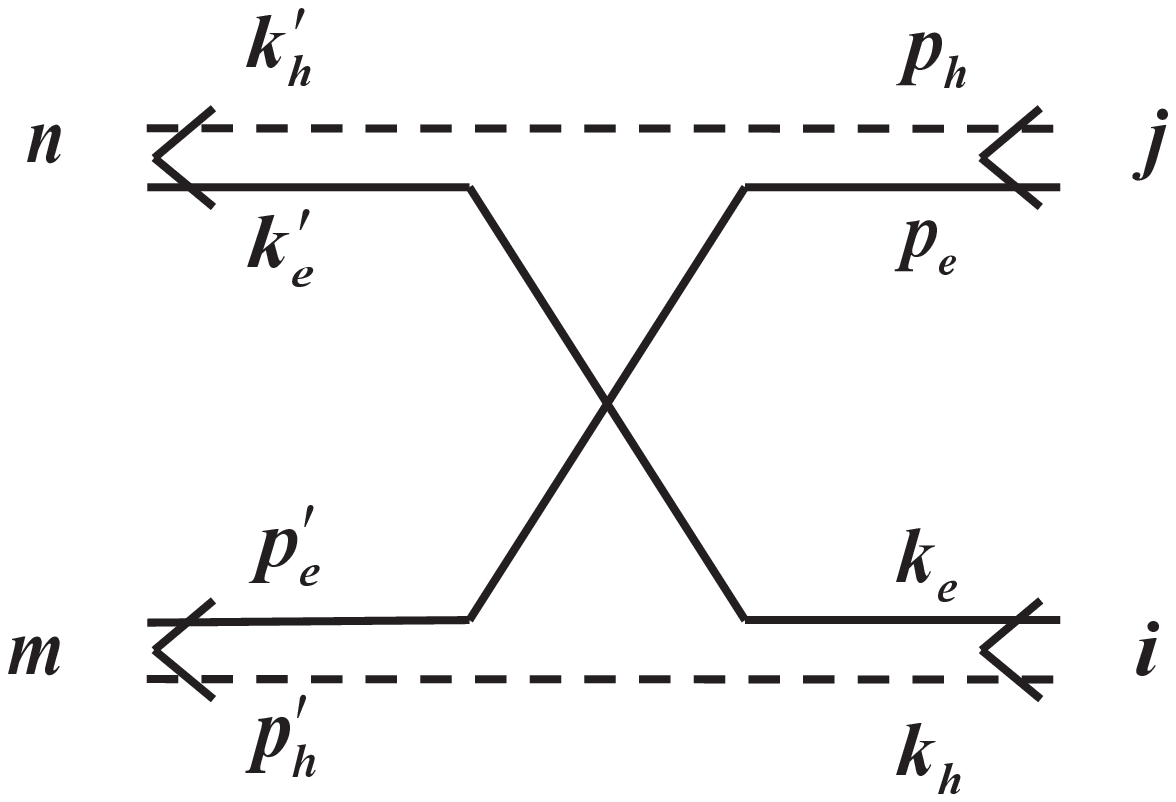}
   \label{fig:fig1b}
 }
\caption{{\small (a) $\lambda_h\big(^{\hspace{.06cm}
n \hspace{.12cm} j\hspace{.05cm}}_{\hspace{.05cm}
m\hspace{.09cm} i\hspace{.05cm}}\big)=\lambda\big(^{\hspace{.06cm}
n \hspace{.12cm} j\hspace{.05cm}}_{\hspace{.05cm}
m\hspace{.09cm} i\hspace{.05cm}}\big)$ for hole exchange, the excitons $m$ and $i$ having the same electron. (b) Pauli scattering $\lambda_e\big(^{\hspace{.06cm}
n \hspace{.12cm} j\hspace{.05cm}}_{\hspace{.05cm}
m\hspace{.09cm} i\hspace{.05cm}}\big)=\lambda\big(^{\hspace{.05cm}
m \hspace{.08cm} j\hspace{.05cm}}_{\hspace{.06cm}
n\hspace{.13cm} i\hspace{.05cm}}\big)$ for electron exchange, the excitons $m$ and $i$ having the same hole.  }}
\label{fig:lambda_eh}
\end{figure}

\section{Biexciton on the exciton basis\label{sec:biX2}}
We consider a system made of two electrons and two holes in a semiconductor: the electrons carry a spin $s=\pm 1/2$ while the holes carry an angular momentum $m$ that we will also call spin. In bulk semiconductors, the hole angular momentum can be $m = (\pm 3/2,\pm 1/2)$, while in narrow quantum wells, it reduces to $m = \pm 3/2$ due to the heavy-light hole energy splitting induced by the well confinement. For simplicity, here we shall neglect the role of light holes and consider heavy holes only. Furthermore, we neglect the warping of the semiconductor valence band and approximate it by a spherical, parabolic band. The usual basis for such a biexciton system is then made of states with two free electrons and two free holes
\bd
a^\dag_{\vk_{e_1}, s_1}a^\dag_{\vk_{e_2} ,s_2}b^\dag_{\vk_{h_1}, m_1}b^\dag_{\vk_{h_2} ,m_2}|v\rangle.\label{eq:free2eh_basis}
\ed
To transform this free-carrier basis into an exciton basis, we make use of the relations which exist between free electron-hole pair creation operators and exciton creation operators, namely,
\bea
B^\dag_{i;s_im_i}&=&\sum_{\vk_e\vk_h}a^\dag_{\vk_{e}, s_i}b^\dag_{\vk_{h}, m_i}\langle \vk_h\vk_e|i\rangle,\label{eq:Btoeh}\\
a^\dag_{\vk_{e}, s_i}b^\dag_{\vk_{h}, m_i}&=&\sum_i B^\dag_{i;s_im_i}\langle i |\vk_e\vk_h\rangle,\label{eq:ehtoB}
\eea
where $|i\ran$ denotes the $i$ exciton state. Using Eq.~(\ref{eq:ehtoB}), we can rewrite the two-free-electron-hole pair states of Eq.~(\ref{eq:free2eh_basis}) in terms of exciton states as
\bd
B^\dag_{i;s_im_i}B^\dag_{j;s_jm_j}|v\rangle,\label{eq:2B_basis}
\ed
with $(s_i,s_j)=(s_1,s_2)$ and $(m_i,m_j)=(m_1,m_2)$. Note that the basis made of $[(s_1,m_1);(s_2,m_2)]$ and $[(s_1,m_2);(s_2,m_1)]$ are equally valid. This means that, for $s_1\ne s_2$ and $m_1\ne m_2$, the basis can be made either of two bright excitons, $(-1/2,3/2)$ and $(1/2,-3/2)$, or of two dark excitons, $(1/2,3/2)$ and $(-1/2,-3/2)$, the bright exciton basis however being more convenient for problems dealing with photons. Note that bright and dark excitons are degenerate if we neglect interband Coulomb processes. \

\begin{figure}[t]
\begin{center}
\includegraphics[trim=3.5cm 7cm 4cm 4cm,clip,width=3.2in] {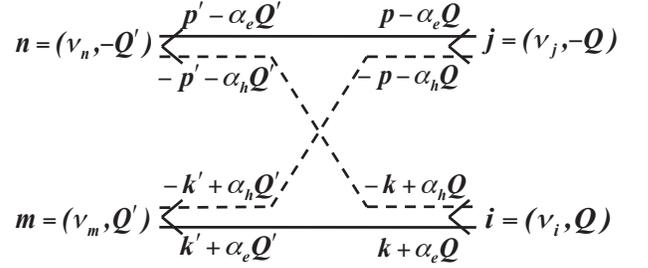}
\end{center}
\caption{{\small Pauli scattering $\lambda\big(^{
(\nu_n,-\vQ^\prime)\hspace{0.1cm} (\nu_j,-\vQ)}_{
\hspace{0.1cm}(\nu_m,\vQ^\prime)\hspace{0.2cm} (\nu_i,\vQ)\hspace{0.15cm}}\big)$ for carrier exchange between an exciton $i=(\nu_i,\vQ)$ and an exciton $j=(\nu_j,-\vQ)$ (see Eq.~(\ref{app:def_lambda})). The exciton $(\nu_i,\vQ)$ is a linear combination of electron-hole pair $(\vk+\alpha_e\vQ,-\vk+\alpha_h\vQ)$ where $\alpha_e=1-\alpha_h=m_e/(m_e+m_h)$  (see Eq.~(\ref{app:B_i_vs_eh})). }}
\label{fig:lambda_h}
\end{figure}

While the great advantage of the exciton basis is to contain part of the electron-hole interaction, actually the strong part leading to exciton bound states, its main disadvantage is to be overcomplete; as a direct consequence, this basis is not orthogonal. It is possible to overcome the difficulties induced by the non-orthogonality of the exciton basis through the commutation technique recently developed for composite boson many-body effects \cite{moniqPhysRep}. The keys of this formalism rely on just four commutators between exciton operators $B_i^\dag$: Fermion exchanges follow from two commutators which read, in the absence of spin degrees of freedom, as
\bea
\left[B_m,B^\dag_i\right]&=&\delta_{mi}-D_{mi},\label{eq:commut_BB}\\
\left[D_{mi},B^\dag_j\right]&=&\sum_n\left[\lambda_h\left(\begin{smallmatrix}
n& j\\ m& i\end{smallmatrix}\right)+\lambda_e\left(\begin{smallmatrix}
n& j\\ m& i\end{smallmatrix}\right)\right]B_i^\dag.\label{eq:commut_DB}
\eea
$D_{mi}$ is called ``deviation-from-boson'' operator because, without it, $B_i^\dag$ would reduce to an elementary boson operator. The Pauli scattering $\lambda_h\left(\begin{smallmatrix}
n& j\\
m& i
\end{smallmatrix}\right)$ corresponds to a hole exchange between excitons $(i,j)$, the excitons $m$ and $i$ having the same electron, as defined in Eq.~(\ref{app:def_lambda_h}) of the appendix and shown in the diagram of Fig.~\ref{fig:fig1a}.  In the same way, the Pauli scattering $\lambda_e\left(\begin{smallmatrix}
n& j\\
m& i
\end{smallmatrix}\right)$ corresponds to an electron exchange, the excitons $m$ and $i$ having the same hole, as defined in Eq.~(\ref{app:def_lambda_e}) and shown in the diagram of Fig.~\ref{fig:fig1b}. $\lambda_h\left(\begin{smallmatrix}
n& j\\
m& i
\end{smallmatrix}\right)$, which is equal to $\lambda_e\left(\begin{smallmatrix}
m& j\\
n& i
\end{smallmatrix}\right)$, is often written as $\lambda\left(\begin{smallmatrix}
n& j\\
m& i
\end{smallmatrix}\right)$ for simplicity.

\begin{figure}[t]
\begin{center}
\includegraphics[trim=2.2cm 3.5cm 3cm 3cm,clip,width=3.4in] {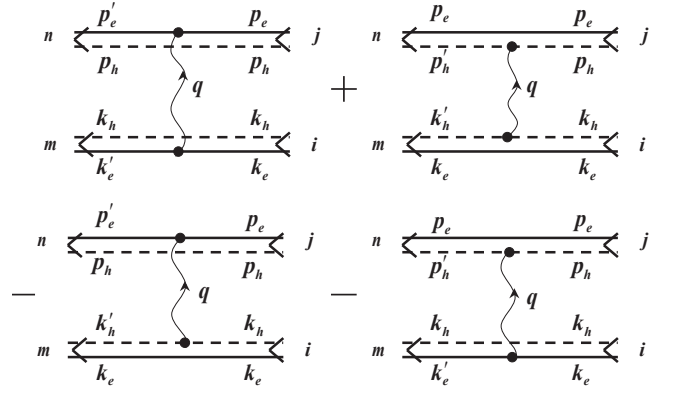}
\end{center}
\caption{{\small Direct Coulomb scattering $\xi^{\rm dir}\big(^{\hspace{.06cm}
n \hspace{.12cm} j\hspace{.05cm}}_{\hspace{.05cm}
m\hspace{.09cm} i\hspace{.05cm}}\big)$ between exciton $i$ and exciton $j$. The ``out'' exciton $m$ is made with the same electron-hole pair as the $i$ exciton. Similarly for excitons $n$ and $j$. This exciton-exciton scattering consists of four terms: two repulsive interactions between electrons and between holes, and two attractive interactions between electron and hole.}}
\label{fig:Xi}
\end{figure}

The other two commutators that handle fermion-fermion interactions, are
\bea
\left[H,B^\dag_i\right]&=&E_iB^\dag_i+V^\dag_i,\label{eq:commut_HB}\\
\left[V^\dag_i,B^\dag_j\right]&=&\sum_{mn}\xi^{\rm dir}\left(\begin{smallmatrix}
n& j\\ m& i\end{smallmatrix}\right)B_m^\dag B_n^\dag.\label{eq:commut_VB}
\eea
The ``creation potential'' $V^\dag_i$ generates, through (\ref{eq:commut_VB}), the direct Coulomb scattering $\xi^{\rm dir}\left(\begin{smallmatrix}
n& j\\ m& i\end{smallmatrix}\right)$ of excitons $i$ and $j$. It consists of four Coulomb processes between the fermionic components of these two excitons: one electron-electron repulsion, one hole-hole repulsion, and two electron-hole attractions, as shown in the diagram of Fig.~\ref{fig:Xi}. The precise expression of the hole-hole part of this scattering is given in Eq.~(\ref{app:def_dirCoul_hh}) of the appendix. \

These four commutators are used to get the biexciton Schr\"{o}dinger equations derived in the next sections. To include the electron and hole degrees of freedom, let us focus on the two relevant cases: \

(a) Carriers with same spins, $s_1=s_2$ and $m_1=m_2$. This corresponds to triplet states for the electron part $(S_e=1,S_e^z=\pm1)$ and triplet-like states for the hole part $(S_h=3,S_h^z=\pm3)$. The associated orbital wave functions then have to be odd with respect to exchange of the two electrons or the two holes in order to fulfill the Pauli exclusion principle.\

(b) Carriers with opposite spins, $s_1=-s_2$ and $m_1=-m_2$. The resulting spin configuration then depends on the way electrons and holes are linearly combined: We can either have a triplet state for the electron part $(S_e=1,S_e^z=0)$ and a triplet-like state for the hole part $(S_h=3,S_h^z=0)$, or a singlet state for the electron part $(S_e=0,S_e^z=0)$ and a singlet-like state for the hole part $(S_h=0,S_h^z=0)$. This case thus requires a more careful derivation since the orbital wave function for triplet state must be odd as in the case (a), but even for the singlet configuration. The biexciton ground state belongs to the set of singlet states.

\section{Biexciton made of electron-hole pairs with same spin $s_1=s_2, m_1=m_2$\label{sec:same_spin}\label{sec:biX3}}

Let us start with triplet biexcitons made of same-spin electrons and same-spin holes and drop the spin indices to make notations of this section lighter. We look for the biexciton eigenstates
\be
(H-\mathcal{E}_\eta)|\Psi^{(\eta)}\rangle=0\label{Psi_eta}
\ee
in the two-free-exciton basis $|ij\rangle=B^\dag_iB^\dag_j|v\rangle$, namely,
\bd
|\Psi^{(\eta)}\rangle=\sum_{ij}\phi_{ij}^{(\eta)}|ij\rangle=\sum_{ij}\phi_{ij}^{(\eta)}B^\dag_iB^\dag_j|v\rangle.\label{eq:Eigst_2Bbs}
\ed
Since $B^\dag_iB^\dag_j=B^\dag_jB^\dag_i$, we can replace the above prefactor by $(\phi_{ij}^{(\eta)}+\phi_{ji}^{(\eta)})/2$. The biexciton state then appears as in Eq.~(\ref{eq:Eigst_2Bbs}) but with the symmetry condition $\phi_{ij}^{(\eta)}=\phi_{ji}^{(\eta)}$.\

\begin{figure}[t]
\begin{center}
\includegraphics[trim=3.5cm 5.5cm 3cm 5cm,clip,width=3.4in] {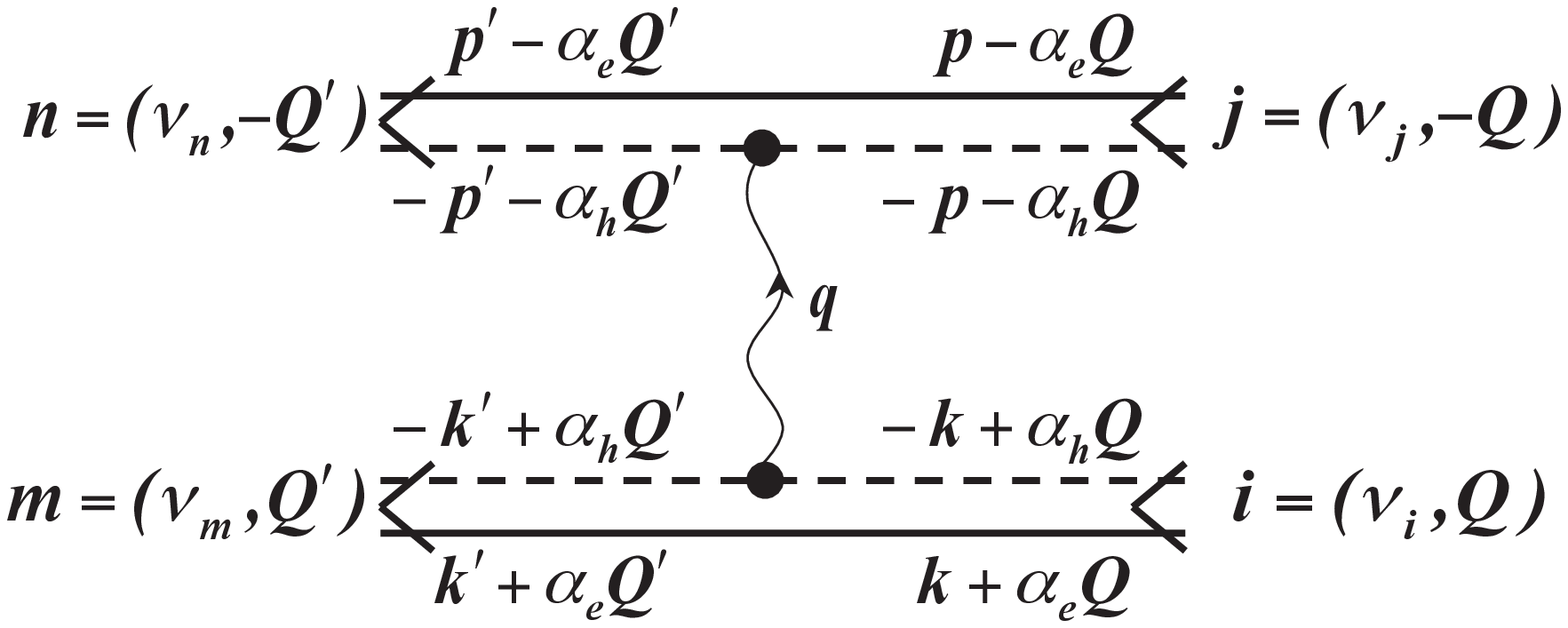}
\end{center}
\caption{{\small Part of the direct Coulomb scattering $\xi^{\rm dir}\big(^{
(\nu_n,-\vQ^\prime)\hspace{0.1cm} (\nu_j,-\vQ)}_{
\hspace{0.1cm}(\nu_m,\vQ^\prime)\hspace{0.2cm} (\nu_i,\vQ)\hspace{0.15cm}}\big)$ between a $i=(\nu_i,\vQ)$ exciton and a $j=(\nu_j,-\vQ)$ exciton coming from hole-hole repulsion (second diagram of Fig.~\ref{fig:Xi}).}}
\label{fig:Xi_part}
\end{figure}

\begin{figure}[t]
\begin{center}
\includegraphics[trim=3cm 4.5cm 3cm 2cm,clip,width=3.4in] {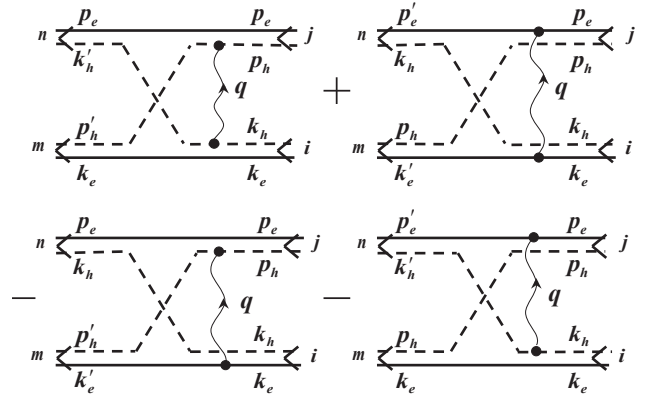}
\end{center}
\caption{{\small ``In'' exchange-Coulomb scattering $\xi^{\rm in}\big(^{\hspace{.06cm}
n \hspace{.12cm} j\hspace{.05cm}}_{\hspace{.05cm}
m\hspace{.09cm} i\hspace{.05cm}}\big)$ between exciton $i$ and exciton $j$. The exciton pair exchanges their holes after Coulomb interactions, while the excitons $m$ and $i$ keep the same electron.}}
\label{fig:Xi_in}
\end{figure}

Equations (\ref{eq:commut_HB}) and (\ref{eq:commut_VB}) allow us to rewrite the biexciton Schr\"{o}dinger equation (\ref{Psi_eta}) as
\bea
0&=&\sum_{ij}\phi_{ij}^{(\eta)}\Big[(E_{ij}-\mathcal{E}_\eta)|ij\rangle+\sum_{rs}\xi^{\rm dir}\left(\begin{smallmatrix}
s& j\\ r& i\end{smallmatrix}\right)|rs\rangle \Big]\nn\\
&=&\sum_{rs}\Big[(E_{rs}-\mathcal{E}_\eta)\phi_{rs}^{(\eta)}+\sum_{ij}\xi^{\rm dir}\left(\begin{smallmatrix}
s& j\\ r& i\end{smallmatrix}\right)\phi_{ij}^{(\eta)}\Big]|rs\rangle\label{eq:Schrod_2Bbs1}
\eea
with $E_{rs}=E_r+E_s$. In the standard case, i.e., when the basis is made of orthogonal states, the above equation forces the bracket to be zero. The situation is more subtle with the exciton basis because, due to Eqs.~(\ref{eq:commut_BB}) and (\ref{eq:commut_DB}), the scalar product of two-exciton states reads as
\bd
\langle v|B_mB_nB^\dag_i B^\dag_j|v\rangle =\Big[\delta_{mi}\delta_{nj}-\lambda\left(\begin{smallmatrix}
n& j\\
m& i
\end{smallmatrix}\right)\Big]+\Big[m\longleftrightarrow n\Big].\label{4Bamtrixelem}
\ed
By projecting Eq.~(\ref{eq:Schrod_2Bbs1}) onto $\langle mn|$, we then find, since $\phi_{ij}^{(\eta)}=\phi_{ji}^{(\eta)}$,
\be
0=(E_{mn}-\mathcal{E}_\eta)\phi_{mn}^{(\eta)}+\sum_{ij}\hat{\xi}^{(\eta)}\left(\begin{smallmatrix}
n& j\\ m& i\end{smallmatrix}\right)\phi_{ij}^{(\eta)},\label{eq:Schrod_2Bbs2}
\ee
where $\hat{\xi}^{(\eta)}\left(\begin{smallmatrix}
n& j\\ m& i\end{smallmatrix}\right)$ is defined as
\be
\hat{\xi}^{(\eta)}\left(\begin{smallmatrix}
n& j\\ m& i\end{smallmatrix}\right)=\xi^{\rm dir}\big(\begin{smallmatrix}
n& j\\ m& i\end{smallmatrix}\big)-\xi^{\rm in}\big(\begin{smallmatrix}
n& j\\ m& i\end{smallmatrix}\big)-\lambda\big(\begin{smallmatrix}
n& j\\ m& i\end{smallmatrix}\big)(E_{ij}-\mathcal{E}_\eta).\label{def:xihat}
\ee
 with  $\xi^{\rm in}\left(\begin{smallmatrix}
n& j\\ m& i\end{smallmatrix}\right)=\sum_{rs}\lambda\left(\begin{smallmatrix}
n& s\\ m& r\end{smallmatrix}\right)\xi^{\rm dir}\left(\begin{smallmatrix}
s& j\\ r& i\end{smallmatrix}\right)$ being ``in'' exchange-Coulomb scattering with Coulomb interactions taking place between the ``in'' exciton pair $(i,j)$, i.e., before hole exchange (see the diagram of Fig.~\ref{fig:Xi_in}).

$\hat{\xi}^{(\eta)}\left(\begin{smallmatrix}
n& j\\ m& i\end{smallmatrix}\right)$  can appear as an effective Coulomb scattering between two excitons although it also depends on the biexciton energy $\mathcal{E}_\eta$. The first two contributions to this effective scattering correspond to the standard combination of Coulomb processes appearing, for example, in the time evolution of two excitons. The third term is more interesting because it only comes from the Pauli exclusion principle. Since the associated Pauli scattering $\lambda\left(\begin{smallmatrix}
n& j\\ m& i\end{smallmatrix}\right)$ is dimensionless, it must go along with an energy to produce an energy-like scattering, this energy actually being an energy difference in order to be gap independent.  \

We now introduce a similar ``out" exchange-Coulomb scattering $\xi^{\rm out}\left(\begin{smallmatrix}
n& j\\ m& i\end{smallmatrix}\right)$, defined as $\xi^{\rm out}\left(\begin{smallmatrix}
n& j\\ m& i\end{smallmatrix}\right)=\sum_{rs}\xi^{\rm dir}\left(\begin{smallmatrix}
 n& s\\ m& r\end{smallmatrix}\right)\lambda\left(\begin{smallmatrix}
s& j\\ r& i\end{smallmatrix}\right)$, Coulomb interactions taking place between the ``out'' exciton pair $(m,n)$ once exchange has occurred. It is easy to check that $\xi^{\rm in}\left(\begin{smallmatrix}
n& j\\ m& i\end{smallmatrix}\right)=\left[\xi^{\rm out}\left(\begin{smallmatrix}
j& n\\ i& m\end{smallmatrix}\right)\right]^*$. Moreover, the ``in'' and ``out'' exchange-Coulomb scatterings are not independent: their difference is related to Pauli scattering via
\bd
\xi^{\rm in}\left(\begin{smallmatrix}
n& j\\ m& i\end{smallmatrix}\right)-\xi^{\rm out}\left(\begin{smallmatrix}
n& j\\ m& i\end{smallmatrix}\right)=(E_{mn}-E_{ij})\lambda\left(\begin{smallmatrix}
n& j\\ m& i\end{smallmatrix}\right),\label{eq:rel_xi_in_out_lambda}
\ed
as easy to recover from calculating the matrix element $\langle v| B_mB_nHB^\dag_iB^\dag_j|v\rangle$ with $H$ acting either on the right or on the left.\

By inserting Eq.~(\ref{eq:rel_xi_in_out_lambda}) into Eq.~(\ref{def:xihat}), we can symmetrize the biexciton Schr\"{o}dinger equation (\ref{eq:Schrod_2Bbs2}) for triplet states as
\bd
0=(E_{mn}-\mathcal{E}_\eta)\phi_{mn}^{(\eta)}+\sum_{ij}\hat\xi_{\rm sym}^{(\eta)}\left(\begin{smallmatrix}
n& j\\ m& i\end{smallmatrix}\right)\phi_{ij}^{(\eta)},\label{eq:Schrod_2Bbs2sym}
\ed
the effective Coulomb scattering, defined as
\bea
\hat\xi_{\rm sym}^{(\eta)}\left(\begin{smallmatrix}
n& j\\ m& i\end{smallmatrix}\right)&=&\xi^{\rm dir}\left(\begin{smallmatrix}
n& j\\ m& i\end{smallmatrix}\right)-\frac{1}{2}\Big[\xi^{\rm in}\left(\begin{smallmatrix}
n& j\\ m& i\end{smallmatrix}\right)+\xi^{\rm out}\left(\begin{smallmatrix}
n& j\\ m& i\end{smallmatrix}\right)\nn\\
&&+\lambda\left(\begin{smallmatrix}
n& j\\ m& i\end{smallmatrix}\right)(E_{mn}+E_{ij}-2\mathcal{E}_\eta)\Big],\label{def_xi_sym1}
\eea
now being symmetrical with respect to the ``in" and ``out" exciton pairs $(i,j)$ and $(m,n)$.
\

\begin{figure}[t]
\begin{center}
\includegraphics[trim=3cm 5.5cm 2.8cm 5cm,clip,width=3.4in] {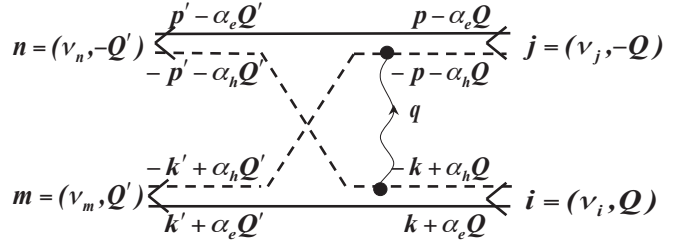}
\end{center}
\caption{{\small Part of the ``in'' exchange-Coulomb scattering $\xi^{\rm in}\big(^{
(\nu_n,-\vQ^\prime)\hspace{0.1cm} (\nu_j,-\vQ)}_{
\hspace{0.1cm}(\nu_m,\vQ^\prime)\hspace{0.2cm} (\nu_i,\vQ)\hspace{0.15cm}}\big)$ between a $i=(\nu_i,\vQ)$ exciton and a $j=(\nu_j,-\vQ)$ exciton, coming from hole-hole repulsion (first diagram of figure~\ref{fig:Xi_in}). The exciton pair exchanges their holes after Coulomb interactions (see Eq.~(\ref{app:inInt1})), the excitons $m$ and $i$ keeping the same electron.}}
\label{fig:Xi_in_hh}
\end{figure}

\section{Biexciton made of electron-hole pairs with opposite spins $s_1=-s_2,~m_1=-m_2$\label{sec:opp_spin}\label{sec:biX4}}
When the electron spins and hole spins are opposite, the derivation of biexciton singlet and triplet states with $S_e^z=S_h^z=0$ requires a more careful analysis of the parity condition induced by the free-carrier fermionic nature. To this end, we first construct biexciton eigenstates in the free-carrier basis and derive the parity condition imposed by the Pauli exclusion principle for singlet and triplet wave functions. We then use Eq.~(\ref{eq:ehtoB}) to rewrite the biexciton eigenstates in terms of exciton operators and rederive the parity condition in the exciton basis. Finally, we write down the biexciton Schr\"{o}dinger equation.

\subsection{Free-carrier basis}
The biexciton eigenstates for opposite electron spins and opposite hole spins can be written on free-carrier states as,
\bea
\lefteqn{|\Psi_{S_e,S_h}^{(\eta)}\rangle=\sum_{\vk_e,\vk^\prime_{e}}\sum_{\vk_{h},\vk^\prime_{h}}\psi^{(\eta,S_e,S_h)}_{\vk_e,\vk^\prime_{e},\vk_{h},\vk^\prime_{h}}}\label{eq:BXeignst_diffspin1}\\
&&\times\Big[a^\dag_{\vk_{e}, +1/2}a^\dag_{\vk^\prime_{e}, -1/2}-(-1)^{S_e}a^\dag_{\vk_{e}, -1/2}a^\dag_{\vk^\prime_{e}, +1/2}\Big]\nn\\
&&\times\Big[b^\dag_{\vk_{h}, +3/2}b^\dag_{\vk^\prime_{h}, -3/2}-(-1)^{S_h}b^\dag_{\vk_{h}, -3/2}b^\dag_{\vk^\prime_{h}, +3/2}\Big]|v\rangle.\nn
\eea
This writing covers singlet state, ($S_e=S_h=0$), as well as ``triplet" state, ($S_e=1,S_h=3$), for which we have a sum instead of a difference of pair states---the biexciton triplet state $S^z=0$ being degenerate with respect to the ones constructed on $(s_1=s_2,~m_1=m_2)$ in Sec.~\ref{sec:biX3}.

Since $a^\dag_{\vk_{e}, -1/2}a^\dag_{\vk^\prime_{e}, +1/2}=-a^\dag_{\vk^\prime_{e}, +1/2}a^\dag_{\vk_{e}, -1/2}$, it is possible to rewrite Eq.~(\ref{eq:BXeignst_diffspin1}) as
\bea
\lefteqn{|\Psi_{S_e,S_h}^{(\eta)}\rangle=\sum_{\vk_e,\vk^\prime_{e}}\sum_{\vk_{h},\vk^\prime_{h}}\hat\psi^{(\eta,S_e,S_h)}_{\vk_e,\vk^\prime_{e},\vk_{h},\vk^\prime_{h}}a^\dag_{\vk_{e}, +1/2}a^\dag_{\vk^\prime_{e}, -1/2}}\label{eq:BXeignst_diffspin2}\\
&&\times\Big[b^\dag_{\vk_{h}, +3/2}b^\dag_{\vk^\prime_{h}, -3/2}-(-1)^{S_h}b^\dag_{\vk_{h}, -3/2}b^\dag_{\vk^\prime_{h}, +3/2}\Big]|v\rangle,\nn
\eea
where $\hat\psi^{(\eta,S_e,S_h)}_{\vk_e,\vk^\prime_{e},\vk_{h},\vk^\prime_{h}}=\psi^{(\eta,S_e,S_h)}_{\vk_e,\vk^\prime_{e},\vk_{h},\vk^\prime_{h}}+(-1)^{S_e}\psi^{(\eta,S_e,S_h)}_{\vk^\prime_e,\vk_{e},\vk_{h},\vk^\prime_{h}}$ follows the parity condition
\bd
\hat\psi^{(\eta,S_e,S_h)}_{\vk_e,\vk^\prime_{e},\vk_{h},\vk^\prime_{h}}=(-1)^{S_e}\hat\psi^{(\eta,S_e,S_h)}_{\vk^\prime_e,\vk_{e},\vk_{h},\vk^\prime_{h}}.
\ed
If we do the same for the hole part, we end up with
\bea
\lefteqn{|\Psi_{S_e,S_h}^{(\eta)}\rangle=\sum_{\vk_e,\vk^\prime_{e}}\sum_{\vk_{h},\vk^\prime_{h}}\phi^{(\eta,S_e,S_h)}_{\vk_e,\vk^\prime_{e},\vk_{h},\vk^\prime_{h}}}\nn\\
&&\times a^\dag_{\vk_{e}, +1/2}a^\dag_{\vk^\prime_{e}, -1/2}b^\dag_{\vk_{h}, +3/2}b^\dag_{\vk^\prime_{h}, -3/2}|v\rangle,\label{eq:BXeignst_diffspin3}
\eea
where $\phi^{(\eta,S_e,S_h)}_{\vk_e,\vk^\prime_{e},\vk_{h},\vk^\prime_{h}}=\hat\psi^{(\eta,S_e,S_h)}_{\vk_e,\vk^\prime_{e},\vk_{h},\vk^\prime_{h}}+(-1)^{S_h}\hat\psi^{(\eta,S_e,S_h)}_{\vk_e,\vk^\prime_{e},\vk^\prime_{h},\vk_{h}}$ has the expected parity condition, namely
\bd
\phi^{(\eta,S_e,S_h)}_{\vk_e,\vk^\prime_{e},\vk_{h},\vk^\prime_{h}}=(-1)^{S_e}\phi^{(\eta,S_e,S_h)}_{\vk^\prime_e,\vk_{e},\vk_{h},\vk^\prime_{h}}=(-1)^{S_h}\phi^{(\eta,S_e,S_h)}_{\vk_e,\vk^\prime_{e},\vk^\prime_{h},\vk_{h}}.
\ed
The biexciton ground state belongs to the set of singlet states $(S_e=S_h=0)$.

\subsection{Exciton basis}

To write the biexciton on the exciton basis, we use Eq.~(\ref{eq:ehtoB}) to rewrite the electron-hole state of Eq.~(\ref{eq:BXeignst_diffspin3}) in terms of bright exciton operators. We find
\be
|\Psi_{S_e,S_h}^{(\eta)}\rangle=\sum_{ij}\phi^{(\eta,S_e,S_h)}_{ij}B^\dag_{i,-1}B^\dag_{j,1}|v\rangle,\label{biXWF1:bright}
\ee
where the prefactor $\phi^{(\eta,S_e,S_h)}_{ij}$, defined as
\be
\phi^{(\eta,S_e,S_h)}_{ij}=\sum_{\vk_e,\vk^\prime_{h}}\sum_{\vk^\prime_{e},\vk_{h}}\langle i|\vk_e\vk^\prime_h\rangle\langle j|\vk^\prime_e\vk_h\rangle\phi^{(\eta,S_e,S_h)}_{\vk_e,\vk^\prime_{e},\vk_{h},\vk^\prime_{h}},\label{eq:def_phi_ij}
\ee
is the biexciton wave function ``in the exciton basis". \

Using $\langle\vk_h\vk_e|\vp_e\vp_h\rangle=\delta_{\vk_e\vp_e}\delta_{\vk_h\vp_h}$ and the exciton closure relation, $\sum_i|i\ran\lan i|=I$,
it is easy to show that from Eqs.~(\ref{eq:def_phi_ij}), (\ref{app:def_lambda_h}), and (\ref{app:def_lambda_e}),  the parity conditions for electron exchange or hole exchange read as
\bea
\phi^{(\eta,S_e,S_h)}_{mn}&=&(-1)^{S_e}\sum_{ij}\lambda_e\left(\begin{smallmatrix}
n& j\\
m& i
\end{smallmatrix}\right)\phi^{(\eta,S_e,S_h)}_{ij}\nn\\
&=&(-1)^{S_h}\sum_{ij}\lambda_h\left(\begin{smallmatrix}
n& j\\
m& i
\end{smallmatrix}\right)\phi^{(\eta,S_e,S_h)}_{ij}.\label{eq:parity_electron_excitonbas}
\eea
By noting that $\lambda_h\left(\begin{smallmatrix}
n& j\\
m& i
\end{smallmatrix}\right)=\lambda_e\left(\begin{smallmatrix}
m& j\\
n& i
\end{smallmatrix}\right)$, these two equations give the parity condition for exciton exchange as
\be
\phi^{(\eta,S_e,S_h)}_{ij}=(-1)^{(S_e+S_h)}\phi^{(\eta,S_e,S_h)}_{ji}.
\ee
\

We now turn to the biexciton Schr\"{o}dinger equation in the exciton basis. We find, from the commutators (\ref{eq:commut_HB}) and (\ref{eq:commut_VB}),
\bea
0&=&(H-\mathcal{E}_\eta^{(S_e,S_h)})|\Psi_{S_e,S_h}^{(\eta)}\rangle\\
&=&\sum_{rs}\Big\{ \left(E_{rs}-\mathcal{E}_\eta^{(S_e,S_h)}\right)\phi^{(\eta,S_e,S_h)}_{rs}\nn\\
&&+\sum_{ij}\xi^{\rm dir}\left(\begin{smallmatrix}
s& j\\
r& i
\end{smallmatrix}\right)\phi^{(\eta,S_e,S_h)}_{ij} \Big\} B^\dag_{r,-1}B^\dag_{s,1}|v\rangle.\nn
\eea
As $\langle v|B_{n,1}B_{m,-1}B^\dag_{r,-1}B^\dag_{s,1}|v\rangle$ reduces to $\delta_{m,r}\delta_{n,s}$ for excitons made of carriers with different spins, the projection of this equation onto $\langle v|B_{n,1}B_{m,-1}$ simply gives
\be
0=(E_{mn}-\mathcal{E}_\eta^{(S_e,S_h)})\phi^{(\eta,S_e,S_h)}_{mn}+\sum_{ij}\xi^{\rm dir}\left(\begin{smallmatrix}
n& j\\
m& i
\end{smallmatrix}\right)\phi^{(\eta,S_e,S_h)}_{ij}.\label{eq:Schrod_diffspin}
\ee
Note that, if we instead project it onto $\langle v|B_{m,1}B_{n,-1}$, we end up with the same equation but with $m$ and $n$ interchanged. So, the resulting Schr\"{o}dinger equation (\ref{eq:Schrod_diffspin}) has to be solved self-consistently with the parity conditions (\ref{eq:parity_electron_excitonbas}),---which is not convenient numerically.\

Moreover, the Schr\"{o}dinger equation (\ref{eq:Schrod_diffspin}) for spin triplet states $(S_e=1,S_h=3)$ does not readily reduce to Eq.~(\ref{eq:Schrod_2Bbs2}), whereas the Schr\"{o}dinger equation must be the same for all triplet states since they are degenerate. To possibly relate these two Schr\"{o}dinger equations and also avoid handling the parity conditions (\ref{eq:parity_electron_excitonbas}), we can introduce two new sets of functions $\varphi^{(c;\eta,S_e,S_h)}_{ij}$ with $c=(e,h)$ and rewrite $\phi^{(\eta,S_e,S_h)}_{ij}$ as
\be
\phi^{(\eta,S_e,S_h)}_{ij}=\varphi^{(c;\eta,S_e,S_h)}_{ij}+(-1)^{S_c}\sum_{mn}\lambda_c\left(\begin{smallmatrix}
j& n\\
i& m
\end{smallmatrix}\right)\varphi^{(c;\eta,S_e,S_h)}_{mn},\label{parity:phi_varphi}
\ee
so that the parity condition (\ref{eq:parity_electron_excitonbas}) is automatically fulfilled whatever $\varphi^{(c;\eta,S_e,S_h)}_{ij}$. By inserting the above equation into Eq.~(\ref{eq:Schrod_diffspin}), we find a Schr\"{o}dinger equation for $\varphi^{(c;\eta,S_e,S_h)}_{ij}$. It reads
\bea
0&=&\left(E_{mn}-\mathcal{E}_\eta^{(S_e,S_h)}\right)\varphi^{(c;\eta,S_e,S_h)}_{mn}\nn\\
&&+\sum_{ij}\Big[\xi^{\rm dir}\left(\begin{smallmatrix}
n& j\\
m& i
\end{smallmatrix}\right)+(-1)^{S_c}\xi^{\rm out}\left(\begin{smallmatrix}
n& j\\
m& i\end{smallmatrix}\right)\label{eq:Schrod_diffspin_2}\\
&&+(-1)^{S_c}\left(E_{mn}-\mathcal{E}_\eta^{(S_e,S_h)}\right)\lambda_c\left(\begin{smallmatrix}
n& j\\
m& i\end{smallmatrix}\right)\Big]\varphi^{(c;\eta,S_e,S_h)}_{ij}.\nn
\eea
If we now use Eq.~(\ref{eq:rel_xi_in_out_lambda}) to rewrite $\xi^{\rm out}\left(\begin{smallmatrix}
n& j\\
m& i\end{smallmatrix}\right)$ in terms of $\xi^{\rm in}\left(\begin{smallmatrix}
n& j\\
m& i\end{smallmatrix}\right)$, the above equation also reads
\bd
0=(E_{mn}-\mathcal{E}_\eta^{(S_e,S_h)})\varphi_{mn}^{(c;\eta,S_e,S_h)}+\sum_{ij}\hat\xi^{(c;\eta)}_{\rm sym}\left(\begin{smallmatrix}
n& j\\
m& i\end{smallmatrix}\right)\varphi_{ij}^{(c;\eta,S_e,S_h)},\label{eq:Schrod_diffspin_sym}
\ed
where the effective scattering now has a symmetrical form with respect to ``in" and ``out" states
\bea
\hat\xi^{(c;\eta)}_{\rm sym}\big(\begin{smallmatrix}
n& j\\
m& i\end{smallmatrix}\big)&=&\xi^{\rm dir}\big(\begin{smallmatrix}
n& j\\
m& i\end{smallmatrix}\big)+\frac{(-1)^{S_c}}{2}\Big[\xi^{\rm in}\big(\begin{smallmatrix}
n& j\\
m& i\end{smallmatrix}\big)+\xi^{\rm out}\big(\begin{smallmatrix}
n& j\\
m& i\end{smallmatrix}\big)\nn\\
&&+\lambda_c\big(\begin{smallmatrix}
n& j\\
m& i\end{smallmatrix}\big)\left(E_{mn}+E_{ij}-2\mathcal{E}_\eta^{(S_e,S_h)}\right)\Big].\label{def_xi_sym2}
\eea
It is then easy to see that, for spin triplet state, $S_h=3$ and $c=h$,  the Schr\"{o}dinger equations (\ref{eq:Schrod_diffspin_sym}) and (\ref{eq:Schrod_2Bbs2}) are indeed identical.  \

Equations (\ref{eq:Schrod_2Bbs2}) and ({\ref{eq:Schrod_diffspin}), in the absence of Coulomb scatterings $\xi^{\rm dir}\left(\begin{smallmatrix}
n& j\\ m& i\end{smallmatrix}\right)$ and Pauli scatterings $\lambda\left(\begin{smallmatrix}
n& j\\ m& i\end{smallmatrix}\right)$, would lead to $\phi_{mn}^{(\eta)}=1$ for $E_{mn}=\mathcal{E}_\eta$ and zero otherwise. The biexciton would then reduce to two free excitons as expected. Interactions between excitons, through both Coulomb and Pauli scatterings, produce a difference between the biexciton energy and the energy of two free excitons. \

By comparing the Schr\"{o}dinger equations (\ref{eq:Schrod_2Bbs2sym}) and (\ref{def_xi_sym1}) for same-spin carriers with the Schr\"{o}dinger equations (\ref{eq:Schrod_diffspin_sym}) and (\ref{def_xi_sym2}) for opposite-spin carriers, we see that there is a sign change in front of the exchange part of the effective scattering (\ref{def_xi_sym2}), depending on singlet and triplet states. This sign change which results from the Pauli exclusion principle, is the reason why two excitons in a singlet state $(S_e=S_h=0)$ can bind together into a molecular state. This can be understood by considering energies close to the energy $2\va_{\nu_0}$ of two ground state excitons labeled by $(\nu_0,\vQ=0)$, which is the expected biexciton energy for temperature much smaller than the energy for exciton internal motion excitation. So, the biexciton binding energy $\delta_\eta=2\va_{\nu_0}-\mathcal{E}_\eta^{(S_e,S_h)}$ is expected to be far smaller than the difference between exciton binding energies $\va_{\nu_i}-\va_{\nu_0}$. We then note that the direct Coulomb scattering $\xi^{\rm dir}\big(\begin{smallmatrix}
 0&  0\\  0&  0\end{smallmatrix}\big)$ is equal to zero; so, $\xi^{\rm dir}$ essentially vanishes for small momentum transfer (see Eq.~(\ref{app:dirInt1})). By contrast, the ``in" exchange-Coulomb scattering is given by \cite{moniqEPJ2003,moniqEP2007}
\begin{equation}
\xi^{\rm in}\big(\begin{smallmatrix}
0& 0\\
0& 0
\end{smallmatrix}\big)= \left\{
\begin{array}{ll}
\displaystyle-\left(8\pi-\frac{315\pi^3}{512}\right)\left(\frac{a_X}{L}\right)^2R_X^{\rm (3D)} & \text{ in 2D }\\
\displaystyle-\frac{26\pi}{3}\left(\frac{a_X}{L}\right)^3R_X^{\rm (3D)} & \text{ in 3D}
\end{array} \right. ,
\end{equation}
with $\xi^{\rm in}\big(\begin{smallmatrix}
0& 0\\
0& 0
\end{smallmatrix}\big)$ equal to $\xi^{\rm out}\big(\begin{smallmatrix}
0& 0\\
0& 0
\end{smallmatrix}\big)$ according to Eq.~(\ref{eq:rel_xi_in_out_lambda}). As a result, the sum of the ``in" and ``out" exchange-Coulomb scatterings, $\xi^{\rm in}\big(\begin{smallmatrix}
0& 0\\
0& 0\end{smallmatrix}\big)+\xi^{\rm out}\big(\begin{smallmatrix}
0& 0\\
0& 0\end{smallmatrix}\big)$, renders the effective scattering $\hat\xi^{(c;\eta)}_{\rm sym}$ overall largely negative for small exciton momenta, while the third term of Eq.~(\ref{def_xi_sym2}), of the order of the biexciton binding energy, is small. This large negative effective scattering allows bound-state solutions to the Schr\"{o}dinger equation (\ref{eq:Schrod_diffspin_sym}). It is then crucial to treat the exchange-Coulomb interactions adequately if one aims at getting reliable results for the ground and excited states. Detailed discussions about the dependence of Coulomb scatterings on electron-to-hole mass ratio and on relative motion momentum of the exciton pair can be found in Refs.~\onlinecite{MoniqPRB2007} and \onlinecite{LauraPRB2010}. For completeness, in \ref{app:sec2}, we have rederived the various scatterings appearing in the biexciton Schr\"{o}dinger equations.\

Equation (\ref{biXWF1:bright}), for $i=(\nu_i,\vk+\vK/2)$ and $j=(\nu_j,-\vk+\vK/2)$, $\vK$ being the biexciton center-of-mass momentum and $\vk$ the relative motion momentum between the exciton pair, allows us to rewrite the biexciton operator in terms of two bright excitons as (see also Eq.~(29) of Ref.~\onlinecite{moniqPRB2009})
\bd
\mathbb{B}_{\eta\vK}^\dag=\sum_{\nu_i\nu_j;\vk}\phi^{(\eta,S_e,S_h)}_\vk(\nu_i,\nu_j) B^\dag_{\nu_i,\vk+\vK/2;-1}B^\dag_{\nu_j,-\vk+\vK/2;1}.\label{BX_operator}
\ed
Since all $\lambda$ and $\xi$ scatterings do not depend on the center-of-mass momentum $\vK$, we can, without any loss of generality, set $\vK=0$ from now on. Such a biexciton is then made of two excitons with opposite momenta. \

Since the parity condition (\ref{eq:parity_electron_excitonbas}) is difficult to numerically implement in the Schr\"{o}dinger equation (\ref{eq:Schrod_diffspin}) for $\phi^{(\eta,S_e,S_h)}_\vk(\nu_i,\nu_j)$, we instead solve a somewhat more complicated equation (\ref{eq:Schrod_diffspin_sym}) in which the parity condition (\ref{parity:phi_varphi}) is already enforced, the function $\varphi^{(c;\eta,S_e,S_h)}_\vk(\nu_i,\nu_j)$ in Eq.~(\ref{parity:phi_varphi}) being a free parameter.\

Let us from now on focus on the biexciton singlet state $(S_e=S_h=0)$ and consider $c=h$ without any loss of generality.
By restricting the exciton level to the ground state $\nu_0$ and by setting $\varphi^{(h;\eta,0,0)}_\vk(\nu_0,\nu_0)=\varphi^{(\eta)}_\vk$, the Schr\"{o}dinger equation (\ref{eq:Schrod_diffspin_sym}) for the singlet state then reduces to
\bea
\lefteqn{-\delta_\eta\Big[\varphi^{(\eta)}_\vk+\sum_{\vk^\prime}\lambda\left(\begin{smallmatrix}
(\nu_0,-\vk)& (\nu_0,-\vk^\prime)\\ (\nu_0,\vk)& (\nu_0,\vk^\prime)\end{smallmatrix}\right)\varphi^{(\eta)}_{\vk^\prime}\Big]}\nn\\
&&\simeq\frac{\vk^2}{M_{X}}\varphi^{(\eta)}_\vk+\sum_{\vk^\prime}\tilde\xi\left(\begin{smallmatrix}
(\nu_0,-\vk)& (\nu_0,-\vk^\prime)\\ (\nu_0,\vk)& (\nu_0,\vk^\prime)\end{smallmatrix}\right)\varphi^{(\eta)}_{\vk^\prime},\label{eq:Schrod_2Bbs4}
\eea
where $M_X=m_e+m_h$ and
\bea
\lefteqn{\tilde\xi\left(\begin{smallmatrix}
(\nu_0,-\vk)& (\nu_0,-\vk^\prime)\\ (\nu_0,\vk)& (\nu_0,\vk^\prime)\end{smallmatrix}\right)=\xi^{\rm dir}\left(\begin{smallmatrix}
(\nu_0,-\vk)& (\nu_0,-\vk^\prime)\\ (\nu_0,\vk)& (\nu_0,\vk^\prime)\end{smallmatrix}\right)}\nn\\
&&+\frac{1}{2}\Big[\xi^{\rm in}\left(\begin{smallmatrix}
(\nu_0,-\vk)& (\nu_0,-\vk^\prime)\\ (\nu_0,\vk)& (\nu_0,\vk^\prime)\end{smallmatrix}\right)+\xi^{\rm out}\left(\begin{smallmatrix}
(\nu_0,-\vk)& (\nu_0,-\vk^\prime)\\ (\nu_0,\vk)& (\nu_0,\vk^\prime)\end{smallmatrix}\right)\nn\\
&&
+\lambda\left(\begin{smallmatrix}
(\nu_0,-\vk)& (\nu_0,-\vk^\prime)\\ (\nu_0,\vk)& (\nu_0,\vk^\prime)\end{smallmatrix}\right)\frac{({\vk'}^2+\vk^2)}{M_X}\Big].
\eea
\

The scatterings $\lambda$ and $\tilde \xi$ depend on $(\vk,\vk^\prime)$ through $|\vk|, |\vk^\prime|$ and the angle $\theta_{\vk\vk^\prime}$ between $\vk$ and $\vk^\prime$, the explicit values of these scatterings being given in \ref{app:sec2}. Since the biexciton singlet state has a zero angular momentum, the $\varphi^{(\eta)}_\vk$ function depends on $|\vk|=k$ only. So, we can first average the various scatterings over the $\theta_{\vk,\vk^\prime}$. Let us call $\lambda(k,k^\prime)$ and $\tilde \xi(k,k^\prime)$ these averaged quantities. We then end up with a 1D integral equation for the function $\varphi^{(\eta)}_k$. It reads as
\bd
-\delta_\eta\Big[\varphi^{(\eta)}_k+\sum_{\vk^\prime}\lambda(k,k^\prime)\varphi^{(\eta)}_{k^\prime}\Big]=\frac{k^2}{M_{X}}\varphi^{(\eta)}_k+\sum_{\vk^\prime}\tilde\xi(k,k^\prime)\varphi^{(\eta)}_{k^\prime}.\label{eq:Schrod_2Bbs5}
\ed
\

This equation is numerically solved to get the binding energies of the biexciton ground and excited states for various hole-to-electron mass ratios. We can also get the $\varphi^{(\eta)}_\vp$ functions, which are related to the biexciton wave functions $\phi^{(\eta)}_\vp$ with proper symmetry via Eq.~(\ref{parity:phi_varphi}). The excited states are expected to mainly come from vibrational modes. To reach rotational modes, it is necessary to include $p$ and $d$ exciton levels. As a consequence, the $\lambda$ and $\xi$ scatterings, as well as the biexciton wave functions $\phi^{(\eta)}_\vk$, would get an angular dependence. The precise treatment of the angular dependence in these scatterings is rather complex and definitely beyond the scope of the present work. A relatively simple, yet nontrivial, biexciton state follows from just considering two ground-state excitons with a nonzero relative-motion angular momentum.\

\begin{figure}[t]
\begin{center}
\epsfig{figure=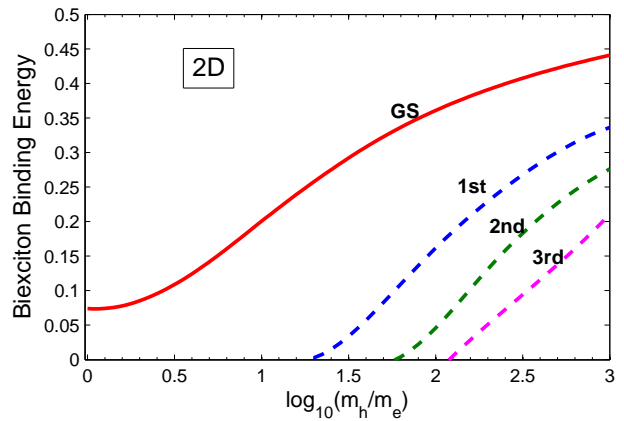,clip=,width=3.6 in}
\end{center}
\caption{(Color online) Binding energies of the 2D
biexciton ground state(GS) and the first three bound excited states in $R_X^{\rm (2D)}=4R_X^{\rm (3D)}$ unit, as a function of the hole-to-electron mass ratio $m_h/m_e$. The ground state binding energy has a minimum for $m_e=m_h$.}
\label{fig:result_bindingEgy2D}
\end{figure}

\begin{figure}[t]
\begin{center}
\epsfig{figure=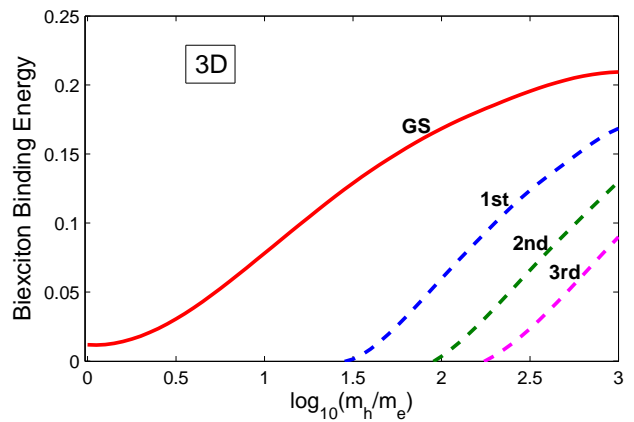,clip=,width=3.6 in}
\end{center}
\caption{(Color online) Same as Fig.~\ref{fig:result_bindingEgy2D} for 3D; the energy unit now is $R_X^{\rm (3D)}$. The curves are qualitatively similar to the 2D curves, though the binding energies are significantly smaller.}
\label{fig:result_bindingEgy3D}
\end{figure}
\section{Results and discussions\label{sec:biX5}}

To solve the Schr\"{o}dinger equation (\ref{eq:Schrod_2Bbs5}) for the biexciton binding energy $\delta_\eta$, we use the normalized exciton ground-state wave functions $\lan \vp|\nu_0\ran$ in 2D and 3D
\bea
\lan \vp|1s\ran^{\rm (2D)}&=& \left(\frac{a_X}{L}\right)\frac{\sqrt{2\pi}}{(1+a_X^2\vp^2/4)^{3/2}},\\
\lan \vp|1s\ran^{\rm (3D)}&=&\left(\frac{a_X}{L}\right)^{3/2}\frac{8\sqrt{\pi}}{(1+a_X^2\vp^2)^2}.
\eea
$L$ is the sample size, $a_X=\hbar^2\epsilon_{sc}/\mu_Xe^2$ is the 3D exciton Bohr radius, with $\mu_X^{-1}=m_e^{-1}+m_h^{-1}$, and $\epsilon_{sc}$ is the static semiconductor dielectric constant, one order of magnitude larger in semiconductor samples than in vacuum.\

The Schr\"{o}dinger equation (\ref{eq:Schrod_2Bbs5}) can be cast into a generalized eigenvalue problem, with matrices spanned by the $\vk$ momentum. To solve it, we sample the $k=|\vk|$ values with $150$ mesh points in 2D and $100$ in 3D, according to $k_i=u_i^3$ where the $u_i$'s are equally distributed, thereby allowing for more sampling in the small $k$ region. The upper cutoff $k_{\rm max}$ (in $a_X^{-1}$ unit) is taken to be $10$ in 3D but $20$ in 2D because the exciton wave function has a larger radial extension (in $\vk$ space) in 2D than in 3D and also because the Coulomb interaction $V_\vq$ decreases as $1/q$ in 2D, i.e., more slowly than the $1/q^2$ dependence it has in 3D (see Eq.~(\ref{app:CoulombPotential_q})). \

\subsection{Biexciton binding energies for ground and excited states}

Solid curves in Figs.~(\ref{fig:result_bindingEgy2D}) and (\ref{fig:result_bindingEgy3D}) show the biexciton binding energies in 2D and 3D as a function of hole-to-electron mass ratio $m_h/m_e$. The results are expressed in terms of the corresponding effective Rydbergs, namely $R_X^{\rm (2D)}$ and $R_X^{\rm (3D)}$, with $R_X^{\rm (2D)}=4R_X^{\rm (3D)}$ and $R_X^{\rm (3D)} = (\mu_X/m_0\epsilon_{sc}^2) 13.6$eV, where $m_0$ is the free electron mass. The curves for the 2D and 3D binding energies are qualitatively similar. However, the values in 2D are significantly larger than those in 3D. This is physically expected since the reduction of dimensionality allows for much stronger Coulomb interactions and more localized wave functions to enhance overlapping. We further notice that both, the 2D and 3D binding energies, have a minimum in the positronium limit, i.e., at $m_h/m_e= 1$; it then increases logarithmically as the mass ratio increases, until it saturates for large mass ratios.\

Our results gives a ground-state biexciton binding energy equal to $0.012R_X^{\rm (3D)}$ for $m_h/m_e=1$ in 3D, which accounts for only about 40\% of the more accurate variational results \cite{Brinkman1972,Akimoto1972}, while it reaches $0.21R_X^{\rm (3D)}$ when $m_h/m_e=1000$, which accounts for about 70\%. In 2D, our calculated binding energies give the ground state at $0.075R_X^{\rm (2D)}$ when $m_h/m_e=1$, which accounts for about 50\% of the best variational result \cite{Kleinman1983}, while it reaches $0.44R_X^{\rm (2D)}$ when $m_h/m_e=1000$, which accounts for about 80\%. All this shows that our approach gives a much better ground-state energy when $m_h/m_e\gg1$, i.e., close to the hydrogen molecule limit, possibly because the exciton wave functions are less deformed when forming a molecule than in the case of lighter hole. \

One important advantage of the present approach over variational procedures is that it allows reaching the biexciton bound and unbound excited states as easily as the ground state. Dashed curves in Figs.~\ref{fig:result_bindingEgy2D} and \ref{fig:result_bindingEgy3D} show the binding energies of the biexciton bound states in 2D and 3D. The number of bound states increases with the mass ratio $m_h/m_e$, this number reducing to 1 for $m_h/m_e\lesssim 20$ in 2D and $m_h/m_e\lesssim30$ in 3D. For a large mass ratio $m_h/m_e=1000$, we find 9 bound states in 2D and 8 in 3D. The binding energy differences become smaller for higher excited states, evidencing a difference in the exciton-exciton interaction compared to the usual harmonic potential which leads to equal energy spacings between eigenstates.   \

\begin{figure}[t]
\begin{center}
\epsfig{figure=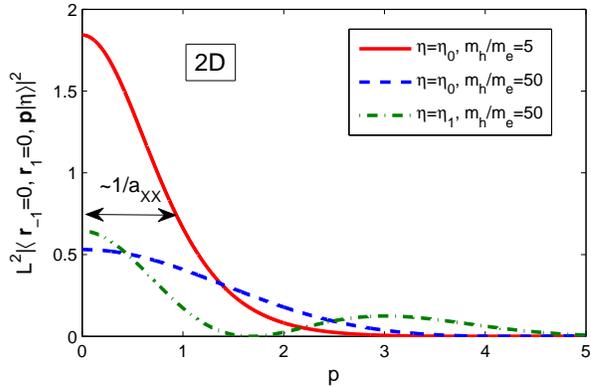,clip=,width=3.4 in}
\end{center}
\caption{(Color online) Plot of the ground state ($\eta=\eta_0$) wave function $L^2|\langle \vr_{-1}=0,\vr_{1}=0, \vp|\eta\rangle|^2$ as a function of $p$ in $a_X^{-1}$ unit, when the mass ratio $m_h/m_e$ is equal to 5 and 50. Its $p$ extension scales as the inverse of the biexciton Bohr radius, $a_{XX}$. We also show the same quantity for the first excited state ($\eta=\eta_1$) when $m_h/m_e=50$. }
\label{fig:WF_r=0peta0}
\end{figure}

\begin{figure}[t]
 \begin{center}
\epsfig{figure=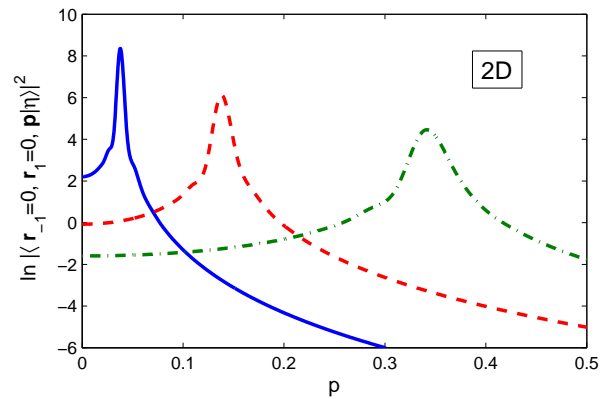,clip=,width=3.4 in}
\end{center}
\caption{(Color online) Plot of $\ln |\langle \vr_{-1}=0,\vr_{1}=0, \vp|\eta\rangle|^2 $ for three unbound biexciton states in a semilog plot, the mass ratio being $m_h/m_e=5$ and the momentum $p$ still in $a_X^{-1}$ unit. Note that the wave function is not here rescaled by $L^2$, so the amplitude of the wave function is significantly larger than for bound states, but its $p$ extension significantly smaller for normalized functions.}
\label{fig:WF_r=0peta3}
\end{figure}

\subsection{Biexciton wave function}
In a previous work on biexciton, we have shown that the wave function of a biexciton made of opposite-spin carriers, with center-of-mass momentum $\vK$, relative motion index $\eta$, and electron and hole total spins $S=(S_e,S_h)$, splits as (see Eq.~(12) in Ref.~\onlinecite{moniqPRB2009})
\be
\lan \vr_{e_1},\vr_{e_2},\vr_{h_1},\vr_{h_2} |\vK,\eta,S\ran=\lan \vR_{XX}|\vK\ran \lan \vr_{-1},\vr_{1},\vu|\eta,S\ran,
\ee
where $\vR_{XX}=(m_e\vr_{e_1}+m_e\vr_{e_2}+m_h\vr_{h_1}+m_h\vr_{h_2})/2(m_e+m_h)$ is the biexciton center-of-mass coordinate, $\vr_{1}=\vr_{e_2}-\vr_{h_1}$ and $\vr_{-1}=\vr_{e_1}-\vr_{h_2}$ are the electron-to-hole distances of the two bright excitons having spins $(\pm1)$, while $\vu=(m_e\vr_{e_2}+m_h\vr_{h_1})/(m_e+m_h)-(m_e\vr_{e_1}+m_h\vr_{h_2})/(m_e+m_h)$ is the distance between the center-of-masses of the two bright excitons.\

For bound biexciton, the wave function $\lan \vr_{-1},\vr_{1},\vu|\eta,S\ran$ has an extension of the order of the exciton size $a_X$ over $\vr_{-1}$ and over $\vr_{1}$, and an extension of the order of the biexciton size $a_{XX}$ over $\vu$. By contrast, for unbound biexciton, the extension over $\vu$ is as large as the sample size $L$, since unbound biexciton resembles very much an exciton with another free exciton moving around anywhere in the sample. Thus, in the case of bound states, dimensional arguments give the normalized relative motion wave functions through
\bea
1&=&\int d\vr_{1} d\vr_{-1}d\vu \left|\lan \vr_{-1},\vr_{1},\vu|\eta,S\ran\right|^2\nn\\
&&\simeq a_X^D a_X^D a_{XX}^D \left|\lan 0,0,0|\eta,S\ran\right|^2,
\eea
which leads to
\be
\left|\lan\vr_{-1}=0,\vr_{1}=0,\vu=0|\eta,S\ran\right|^2\simeq\left(\frac{1}{a_X^2a_{XX}}\right)^D,
\ee
while, in the case of unbound states, $a_{XX}$ is replaced by $L$; so,
\be
\left|\lan\vr_{-1}=0,\vr_{1}=0,\vu|\eta,S\ran\right|^2\simeq\left(\frac{1}{a_X^2L}\right)^D.
\ee
\

To obtain $\lan \vr_{-1},\vr_{1},\vp|\eta,S\ran$, which is of physical relevance in photon absorption, we perform a Fourier transform as
\be
\lan \vr_{-1},\vr_{1},\vp|\eta,S\ran=\int d\vu\lan\vp|\vu\ran\lan \vr_{-1},\vr_{1},\vu|\eta,S\ran,\label{eq:wf_FT:up}
\ee
where $\lan\vp|\vu\ran=e^{i\vp\cdot \vu}/L^{D/2}$; so, the extension over $\vp$ of $\lan \vr_{-1},\vr_{1},\vp|\eta,S\ran$ is of the order of $1/a_{XX}$ for bound states (see Fig.~\ref{fig:WF_r=0peta0}), and of the order of $1/L$ for unbound states (see Fig.~\ref{fig:WF_r=0peta3}). The same dimensional arguments then give, in the case of bound states,
\be
\left|\lan \vr_{-1}=0,\vr_{1}=0,\vp=0|\eta,S\ran\right|^2\simeq\left(\frac{a_{XX}}{a_X^2L}\right)^D,\label{WF:DimArg:rrp}
\ee
and, in the case of unbound states,
\be
\left|\lan \vr_{-1}=0,\vr_{1}=0,\vp|\eta,S\ran\right|^2\simeq\left(\frac{1}{a_X^2}\right)^D.
\ee
\

We can compute $\lan \nu_i,\nu_j,\vp|\eta,S\ran$ from $\lan \vr_{-1},\vr_{1},\vp|\eta,S\ran$ through a double Fourier transform ``in the exciton sense" (see also Eq.~(31) of Ref.~\onlinecite{moniqPRB2009}), namely,
\be
\lan \nu_i,\nu_j,\vp|\eta,S\ran=\int d\vr_{-1}d\vr_{1}\lan\nu_i|\vr_{-1}\ran \lan\nu_j|\vr_{1}\ran \lan \vr_{-1},\vr_{1},\vp|\eta,S\ran,\label{eq:wf_DFT:nunu}
\ee
which also reads as
\be
 \lan \vr_{-1},\vr_{1},\vp|\eta,S\ran=\sum_{\nu_i,\nu_j}\lan\vr_{-1}|\nu_i\ran \lan\vr_{1}|\nu_j\ran\lan \nu_i,\nu_j,\vp|\eta,S\ran.\label{eq:wf_DFT:rr}
\ee
Note that for $\vr_{1}$ or $\vr_{-1}$ equal to 0, the $\nu$ exciton levels that survive in the above $\nu$ sum are $s$-like states only. So, the $\lan \nu_i,\nu_j,\vp|\eta,S\ran$ function just is the function $\phi^{(\eta,S)}_\vp(\nu_i,\nu_j)$ given in Eq.~(\ref{BX_operator}). Since, when numerically solving the biexciton Schr\"{o}dinger equation (\ref{eq:Schrod_2Bbs5}) for singlet state ($S=0$), we have restricted the sum over $\nu$ to the ground state $\nu_0$, we must for consistency also keep $\nu_0$ only in the $\nu$ sum of Eq.~(\ref{eq:wf_DFT:rr}).\

Figure \ref{fig:WF_r=0peta0} shows $|\lan \vr_{-1}=0,\vr_{1}=0,\vp|\eta,S=0\ran|^2$ for the ground state when the mass ratio is $m_h/m_e=5$, and for two bound states when the mass ratio is $m_h/m_e=50$. Note that Eq.~(\ref{WF:DimArg:rrp}) forces us to plot bound-state wave functions through $L^2|\lan \vr_{-1}=0,\vr_{1}=0,\vp|\eta,S=0\ran|^2$ in order to have a quantity independent of sample size $L$. For unbound states, the $\left|\lan \vr_{-1}=0,\vr_{1}=0,\vp|\eta,S=0\ran\right|^2$ function is peaked on momenta $\vp_\eta$ which depend on the unbound biexciton energies. Figure \ref{fig:WF_r=0peta3} shows that this peaked function broadens when the unbound biexciton energy increases, the broadening being due to exciton-exciton interactions.

\

\subsection{Biexciton absorption spectrum in pump-probe experiment}

(i) Let us now first consider an initial state made of one circularly polarized photon $\sigma_+$ with momentum $\vQ_{ph}$ and frequency $\omega$, and one exciton already present in the sample, this $(\nu_0,\vQ_i)$ exciton having an opposite circular polarization, $\sigma_-$. After photon absorption, the final state contains two electron-hole pairs, their center-of-mass momentum being $\vK_i=\vQ_{ph}+\vQ_i$. The photocreated exciton interacts with the exciton present in the sample to possibly form a biexciton. Since we are mainly interested in low-lying biexciton states, we shall focus on singlet states $(S=0)$. The Fermi golden rule gives the photon absorption as $(-2)$ times the imaginary part of the response function to one photon $(\omega,\vQ_{ph})$. This response function reads as (see Eq.~(36) of Ref.~\onlinecite{moniqPRB2009})
\bea
\lefteqn{S_{XX}(\omega,\vQ_{ph};\vQ_i)=}\nn\\
&&\sum_\eta \frac{f^{(\eta)}_{XX}(\vp_i)}{\omega+E_{\nu_0,\vQ_i}-\left[\mathcal{E}_\eta+\frac{(\vQ_{ph}+\vQ_i)^2}{4M_X}\right]+i0^+},\label{BX:repsfun}
\eea
where $\vp_i=(\vQ_i-\vQ_{ph})/2$ is the relative motion momentum of the $(X,X)$ pair and $E_{\nu_0,\vQ_i}$ is the free exciton energy given by Eq.~(\ref{app:E_iQ}). The biexciton oscillator strength $f^{(\eta)}_{XX}(\vp)$ in Eq.~(\ref{BX:repsfun}) is given by
\bea
f^{(\eta)}_{XX}(\vp)&=&|\Omega|^2L^D\Big|\sum_\nu\lan\vr=0 |\nu\ran \lan\nu_0,\nu,\vp|\eta\ran\Big|^2\nn\\
&=&|\Omega|^2L^D\Big|\lan\nu_0,\vr=0,\vp|\eta\ran\Big|^2,\label{OS_BX}
\eea
where $\Omega$ is the vacuum Rabi coupling, and $ \lan\nu_i,\nu_j,\vp|\eta\ran$ is the singlet biexciton wave function $ \lan\nu_i,\nu_j,\vp|\eta,S=0\ran$. \

Using Eq.~(\ref{WF:DimArg:rrp}), we can show that the biexciton oscillator strength $f^{(\eta,S=0)}_{XX}(\vp)$ for $\vp$ close to zero, is related to the exciton oscillator strength
\be
f_X=|\Omega|^2L^D|\lan\vr=0 |\nu\ran|^2\simeq|\Omega|^2(L/a_X)^D
\ee
via\cite{moniqPRB2009}
\be
f^{(\eta,S=0)}_{XX}\simeq (a_{XX}/L)^Df_X .\label{OS_bX_ratio}
\ee
The prefactor $(a_{XX}/L)^D$ corresponds to the localization into a biexciton volume $a^D_{XX}$ of the exciton present in the sample and initially delocalized over a volume $L^D$. For very large sample size $L$, this {\it a priori} prevents using the same scale to draw bound and unbound biexciton absorption spectra through linear response to a photon field.\

(ii) We now turn to a more complicated situation in which one circularly polarized photon, $\sigma_+$, is absorbed in a dilute exciton gas having $N_X$ excitons with opposite circular polarization, $\sigma_-$, which is what currently happens in pump-probe experiments: One first prepares a dilute exciton gas using a circularly polarized photon beam with low pump power; then we probe this gas with a weak photon beam having opposite polarization. We assume that these $N_X$ excitons all are in the exciton ground state $\nu_0$, the temperature being too small to have excited exciton states populated. For a dilute exciton gas, i.e., $N_X(a_X/L)^D\ll 1$, we may neglect exciton many-body effects, since these effects scale at least quadratically in the exciton density $n_X(=N_X/L^D)$. Indeed, the Pauli scattering for fermion exchanges between {\it two} excitons leads to terms in $n_X^2$. Such many-body effects would alter the absorption spectra presented below, because they affect the exciton energy states, and accordingly the biexciton state. In addition, the photocreated biexciton can interact with other excitons in the exciton gas. As a first approximation, we consider the dilute exciton gas as a set of noninteracting classical particles. The $\vQ_i$ exciton distribution for finite temperature $T$ then is just the Boltzmann distribution
\be
N(\vQ_i,T)=\frac{N_X}{L^D}\left(\frac{2\pi}{M_Xk_BT}\right)^{D/2}e^{-\vQ_i^2/2M_Xk_BT},\label{eq:N_kT}
\ee
normalized through $\sum_{\vQ_i}N(\vQ_i,T)=N_X$. \

Since $\vQ_{ph}\approx0$ on the characteristic electron scale, we can write the response function of $N_{ph}$ photons to a $N(\vQ_i,T)$ exciton distribution as
\be
S^{(XX)}(\omega,T)=N_{ph}\sum_{\vQ_i}N(\vQ_i,T)S_{XX}(\omega,0;\vQ_i),\label{SI0}
\ee
with $S_{XX}(\omega,\vQ_{ph};\vQ_i)$ given in (\ref{BX:repsfun}).

\begin{figure}[t]
\begin{center}
\epsfig{figure=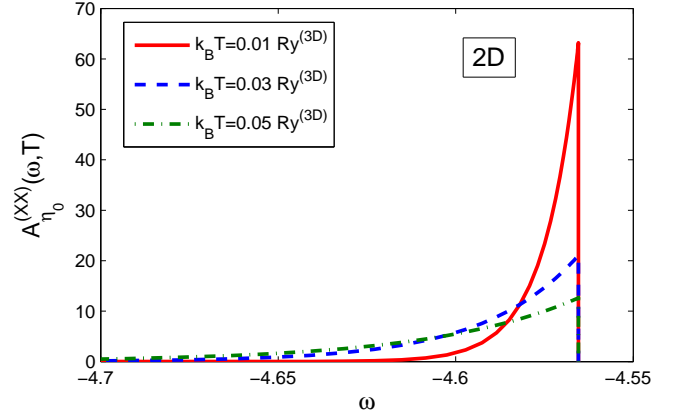,clip=,width=3.6 in}
\end{center}
\caption{(Color online) Absorption amplitude $A^{(XX)}_{\eta_0}(\omega,T)$, defined in Eq.~(\ref{SI2}), for the biexciton ground state in 2D, as a function of the photon energy $\omega$ in $R_X^{\rm (3D)}$ unit for various temperatures $T$. We find an asymmetric low-energy peak at $-4.57R_X^{\rm (3D)}$, which is associated with the biexciton ground state having a binding energy $0.57R_X^{\rm (3D)}$ below $R_X^{\rm (2D)}=4R_X^{\rm (3D)}$. The hole-to-electron mass ratio is $m_h/m_e=5$, as in usual GaAs samples, and $8\pi|\Omega|^2N_{ph}N_X\rho/L^D=1$ is set equal to 1, the $L^D$ factor coming from the wave function part of this equation, due to Eq.~(\ref{WF:DimArg:rrp}). }
\label{fig:result_Abs_bound}
\end{figure}

\subsubsection{Bound biexciton}

By replacing the $\vQ_i$ sum in Eq.~(\ref{SI0}) by an integral with a constant density of states $\rho$ for 2D systems, and by setting $\va=\vQ_i^2/2M_X$, we find that the absorption spectrum associated with a biexciton bound state $\eta_0$ is given by
\bea
A_{\eta_0}^{(XX)}(\omega,T)&=&\frac{4\pi^2|\Omega|^2N_{ph}N_X}{M_Xk_BT}\int\rho d\va e^{-\va/k_BT}\nn\\
&&\times\Big|\lan\nu_0,\vr=0 ,p=\sqrt{M_X\va/2}|\eta_0\ran\Big|^2\nn\\
&&\times \delta\left(\omega+\va_{\nu_0}-\mathcal{E}_{\eta_0}+\va/2\right).\label{SI1}
\eea
For a photon detuning $\delta^{(\eta_0)}_{ph}=\mathcal{E}_{\eta_0}-\va_{\nu_0}-\omega>0$, this leads to
\bea
A_{\eta_0}^{(XX)}(\omega,T)&=&\frac{8\pi^2|\Omega|^2N_{ph}N_X\rho}{M_Xk_BT} e^{-2\delta^{(\eta_0)}_{ph}/k_BT}\label{SI2}\\
&&\times\Big| \lan\nu_0,\vr=0,p=\sqrt{M_X\delta^{(\eta_0)}_{ph}}|\eta_0\ran\Big|^2.\nn
\eea
\

Figure \ref{fig:result_Abs_bound} shows the absorption $A^{(XX)}_{\eta_0}(\omega,T)$ (\ref{SI2}) associated with the biexciton ground state $\eta_0$ in the presence of a dilute exciton gas as a function of the photon energy $\omega$ for various temperatures. We identify the biexciton ground state with the low-energy peak lying $0.57R_X^{\rm (3D)}$---which is the biexciton binding energy---below a larger peak at $-4R_X^{\rm (3D)}$, shown in Fig.~\ref{fig:result_Abs_unbound}, that accounts for the photocreated exciton scattered by the free excitons present in the sample. Note the $(L/a_X)^D$ factor being the scale between the two figures. The lineshape of the ground-state biexciton peak is asymmetric with a long tail on the lower-energy side, the peak strength decreasing with increasing temperature. This low-energy tail reflects the fact that the photon energy required to meet the biexciton ground-state energy, as a result of energy conservation in Eq.~(\ref{SI1}), is smaller if the exciton already present in the sample has a larger kinetic energy. So, the biexciton binding energy corresponds to the upper sharp edge of the peak, since the tail comes from biexciton with larger relative motion momenta. Of course, this upper sharp edge is smoothed into a high-energy tail if we include a broadening of the delta function in Eq.~(\ref{SI1}), which amounts to replacing $i0^+$ in Eq.~(\ref{BX:repsfun}) by a finite width $i\gamma$. This low-energy tail is also found in the photoluminescence spectra of 1D quantum wires \cite{Crottini2002}.\

\subsubsection{Unbound biexciton}

Since the unbound biexciton energies are close to the exciton energies, photoabsorption with formation of an unbound biexciton $A^{(XX)}$ is mixed with photoabsorption with formation of an exciton $A^{(X)}$, depending on the capture rate $f$ of an exciton by the photocreated exciton. As a first approximation, one can write the resulting absorption spectrum as
\be
fA^{(XX)}+(1-f)A^{(X)}=A^{(X)}+f\left[A^{(XX)}-A^{(X)}\right].\label{def:Abs_GS3}
\ee
For $N_X$ excitons in a sample volume $L^D$, the capture rate $f$ should be of the order of the exciton volume divided by the average volume occupied by one free exciton in the sample, namely,
 \be
 f\simeq \frac{a_X^D}{L^D/N_X}=N_X\left(\frac{a_X}{L}\right)^D.
 \ee
However, since excitons with different $\vQ_i$'s contribute differently to the biexciton absorption, as seen from Eq.~(\ref{SI0}), the absorption spectrum should in fact read, instead of Eq.~(\ref{def:Abs_GS3}), as
\bea
A(\omega,T)&=&A^{(X)}(\omega)+\left(\frac{a_X}{L}\right)^D\sum_{\vQ_i}N(\vQ_i,T)\nn\\
&&\times\left[ A^{(XX)}(\omega,\vQ_i)-A^{(X)}(\omega)\right],\label{def:Abs_GS4}
\eea
with $N(\vQ_i,T)$ given by Eq.~(\ref{eq:N_kT}). We see that the absorption spectrum reduces to the exciton spectrum in the absence of free excitons, $N_X=0$, as physically required.\

The exciton absorption spectrum $A^{(X)}(\omega)$ is made of delta peaks centered on the exciton energies $\va^{(\nu)}$, and weighted by the value at $\vr=0$ of the exciton wave function squared $|\lan \vr=0|\nu\ran|^2$. When taking into account the finite exciton lifetime, these delta peaks broaden into Lorentzian functions with finite width.\

Although, according to Eq.~(\ref{def:Abs_GS4}), photon absorption contains an exciton and a biexciton part, we have chosen to only show here the part of the spectrum coming from unbound biexcitons, namely,
\bea
A^{(XX)}(\omega,T)&=&
\frac{4\pi^2|\Omega|^2N_{ph}N_X}{M_Xk_BT}\sum_{\eta}\int\rho d\va e^{-\va/k_BT}\nn\\
&&\times\Big|\lan\nu_0,\vr=0 ,\sqrt{M_X\va/2}|\eta\ran\Big|^2\nn\\
&&\times \delta\left(\omega+\va_{\nu_0}-\mathcal{E}_\eta+\va/2\right),\label{SI3}
\eea
in order to avoid ambiguity coming from the relative weights of the exciton and biexciton lifetimes, which are sample dependent. To compute $A^{(XX)}(\omega,T)$, we replace the discrete sum over $\eta$ by a continuous sum over $\vp_\eta$ and we introduce a finite lifetime by replacing the delta function by a Lorentzian function having a small half-width $\gamma$, namely,
\be
\delta(\omega)\rightarrow \frac{\gamma/\pi}{\omega^2+\gamma^2}.\label{deltaf}
\ee
\
\begin{figure}[t]
\begin{center}
\epsfig{figure=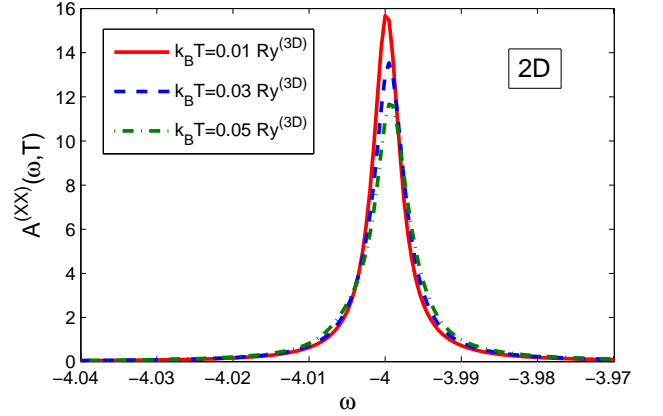,clip=,width=3.6 in}
\end{center}
\caption{(Color online) Absorption amplitude $A^{(XX)}(\omega,T)$, defined in Eq.~(\ref{SI3}), for the biexciton unbound states in 2D, as a function of the photon energy $\omega$ in $R_X^{\rm (3D)}$ unit for various temperatures. We find a large peak centered on $-R_X^{\rm (2D)}=-4R_X^{\rm (3D)}$. We have taken an hole-to-electron mass ratio $m_h/m_e=5$, a broadening $\gamma=0.002R_X^{\rm (3D)}$ in Eq.~(\ref{deltaf}), and we have set $8\pi|\Omega|^2N_{ph}N_X\rho/a_X^D=1$. }
\label{fig:result_Abs_unbound}
\end{figure}

Figure \ref{fig:result_Abs_unbound} shows the lineshape of the peak located at $-4R_X^{\rm (3D)}$, which is the $1s$ exciton level in 2D quantum wells. This peak corresponds to unbound biexcitons, i.e., exciton-exciton scattering states. We see that this peak spreads on both sides of the $1s$ exciton level, due to energy conservation enforced by the broadened delta function. The peak lineshape essentially is Lorentzian with a peak height slightly decreasing with temperature. This can be attributed to broadening at large relative motion momentum induced by exciton-exciton scatterings because contributions from large relative momenta start to weight in when the temperature gets large. Due to energy conservation enforced by the delta function in Eq.~(\ref{SI3}), the broadening induced by temperature may lead to unbound states with a much broader lineshape on top of the sharp exciton peak. The broadening of the peak remains close to $0.01R_X^{\rm (3D)}$ even though the temperature changes from $0.01R_X^{\rm (3D)}$ to $0.03R_X^{\rm (3D)}$. This indicates a ``non-thermal broadening" behavior, since, for such temperatures, the intrinsic broadening due to exciton-exciton scattering is dominant over thermal broadening. \

\section{Conclusions}
We have constructed the Schr\"{o}dinger equation for biexciton written in the exciton basis, instead of the standard free-carrier basis. To choose such a basis is motivated by the physical fact that Coulomb interaction between electron and hole is far stronger than between two dipole-like excitons. However, in this exciton formulation, we must handle the fact that two excitons can exchange their carriers. The exact handling of this carrier exchange, as induced by the Pauli exclusion principle, is made possible thanks to the recently developed composite boson many-body formalism. \

By restricting the exciton levels to the ground state, we have numerically solved the resulting biexciton Schr\"{o}dinger equation for the ground state, as well as for the bound and unbound excited states in 2D and 3D systems. Our results for the ground-state binding energies agree reasonably well with those obtained from variational methods. The main advantage of the present approach is to easily reach excited states, which are out of reach from usual variational procedures. We then use the obtained biexciton wave functions to calculate the optical absorption spectrum in the presence of a dilute exciton gas in 2D. The spectrum shows a small asymmetric low-energy peak associated with the biexciton ground state, and larger peaks coming from exciton-exciton scattering states.

\section{Acknowledgment}

This work is supported by the National Science Council of Taiwan under Contract No. NSC 101-2112-M-001-024-MY3 and Academia Sinica, Taiwan. M.C. wishes to thank the Research Center for Applied Sciences of the Academia Sinica in Taiwan for various invitations. We also acknowledge fruitful discussions with K. V. P. Latha.

\renewcommand{\thesection}{\mbox{Appendix~\Roman{section}}} 
\setcounter{section}{0}

\renewcommand{\theequation}{\mbox{A.\arabic{equation}}} 
\setcounter{equation}{0} %
\section{Exciton operator in momentum space\label{app:sec1}}

The index $i$ in a state $|i\ran$ of an exciton made of electron-hole pairs in a translationally invariant system, refers to the pair center-of-mass momentum $\vQ_i$ and the internal motion index of the exciton state at hand, $\nu_i$, i.e., $i=(\nu_i,\vQ_i)$. The wave function of this $(\nu_i,\vQ_i)$ exciton splits as
\be
\phi_i(\vr_e,\vr_h)=\left\lan \vR|\vQ_i\right\ran\lan \vr|\nu_i\ran,
\ee
where $\vR=(m_e \vr_e+m_h \vr_h)/(m_e+m_h)$ is the center-of-mass coordinate of the pair and $ \vr=\vr_e-\vr_h$ is the distance between electron and hole. The center-of-mass momentum wave function is a plane wave, $\left\lan \vR|\vQ_i\right\ran=e^{i\vQ_i\cdot\vR}/L^{D/2}$ for a size $L$ sample. By writing the relative motion wave function as $\lan \vr|\nu\ran=\sum_{\vp_i}\lan \vr|\vp_i\ran\lan\vp_i|\nu\ran $, we can rewrite $\phi_i(\vr_e,\vr_h)$ as
\be
\phi_i(\vr_e,\vr_h)=\sum_{\vp_i} \lan\vp_i|\nu_i\ran\frac{e^{i(\vp_i+\alpha_e \vQ_i)\cdot \vr_e}}{L^{D/2}}\frac{e^{i(-\vp_i+\alpha_h \vQ_i)\cdot \vr_h}}{L^{D/2}},\label{eq:func_phi}
\ee
where we have set
\be
\alpha_e=\frac{m_e}{m_e+m_h}=1-\alpha_h.
\ee
So, the electron and hole momenta read in terms of the center-of-mass and relative motion momenta of the pair, $\vQ_i$ and $\vp_i$, as
\be
\vk_e=\vp_i+\alpha_e\vQ_i,\qquad \vk_h=-\vp_i+\alpha_h\vQ_i.
\ee
\

The exciton relative motion wave function $\lan\vk|\nu_i\ran$ obeys the Schr\"{o}dinger equation
\be
\left(\frac{\vk^2}{2\mu_X}-\va_{\nu_i}\right) \lan\vk|\nu_i\ran-\sum_{\vq\not=0}V_\vq \lan\vk+\vq|\nu_i\ran=0,\label{app:X_schrodeq}
\ee
where $\mu_X^{-1}=m_e^{-1}+m_h^{-1}$ is the electron-hole pair reduced mass, while the Coulomb potential $V_\vq$ is given by
\begin{equation}
\label{app:CoulombPotential_q}
V_\vq = \left\{
\begin{array}{rl}
2\pi e^2/\epsilon_{sc}L^2q & \text{ in 2D }\\
4\pi e^2/\epsilon_{sc}L^3q^2 & \text{ in 3D}
\end{array} \right. .
\end{equation}
\

From Eq.~(\ref{eq:func_phi}), we can deduce the creation operator of the $i$ exciton as
\be
B^\dag_i=B^\dag_{\nu_i,\vQ_i}=\sum_{\vk}\lan\vk|\nu_i\ran a^\dag_{\vk+\alpha_e \vQ_i}b^\dag_{-\vk+\alpha_h \vQ_i},\label{app:B_i_vs_eh}
\ee
This operator creates a single-pair eigenstate
\be
(H-E_i)B^\dag_i|v\ran=0.
\ee
The exciton energy $E_i$ contains a relative motion part and a center-of-mass kinetic energy, namely,
\be
E_i=E_{\nu_i,\vQ_i}=\va_{\nu_i}+\frac{\vQ_i^2}{2M_X},\label{app:E_iQ}
\ee
where $M_X=m_e+m_h$.

\renewcommand{\theequation}{\mbox{B.\arabic{equation}}} 
\setcounter{equation}{0} 

\section{On the various scatterings of two excitons\label{app:sec2}}
In this appendix, we rederive various scatterings appearing in the biexciton Schr\"{o}dinger equation \cite{moniqPhysRep,MoniqPRB2007}. Without any loss of generality in calculating these scatterings, we set the biexciton center-of-mass momentum $\vK$ to be zero. The two excitons then have opposite momenta. This just amounts to working in the reference frame of the center-of-mass of the exciton pair.

\subsection{Pauli scattering}

Let us first consider the Pauli scattering for hole exchange between excitons starting in ``in" states $(i,j)$ and ending in ``out" states $(m,n)$, the excitons $m$ and $i$ having the same electron, as shown in the diagram of Fig.~\ref{fig:fig1a}. This scattering is given, as read from the figure, by
\be
\lambda_h\left(\begin{smallmatrix}
n& j\\
m& i
\end{smallmatrix}\right)=\sum\langle m|\vk'_e\vk'_h\rangle\langle n|\vp'_e\vp'_h\rangle\langle\vp_h\vp_e|j\rangle\langle\vk_h\vk_e|i\rangle,\label{app:def_lambda_h}
\ee
Similarly, the Pauli scattering for electron exchange shown in the diagram of Fig.~\ref{fig:fig1b}, with the excitons $m$ and $i$ having the same hole, is given by
\be
\lambda_e\left(\begin{smallmatrix}
n& j\\
m& i
\end{smallmatrix}\right)=\sum\langle m|\vp'_e\vp'_h\rangle\langle n|\vk'_e\vk'_h\rangle\langle\vp_h\vp_e|j\rangle\langle\vk_h\vk_e|i\rangle.\label{app:def_lambda_e}
\ee
It is easy to check that $\lambda_e\left(\begin{smallmatrix}
n& j\\
m& i
\end{smallmatrix}\right)=\lambda_h\left(\begin{smallmatrix}
m& j\\
n& i
\end{smallmatrix}\right)$, the Pauli scattering $\lambda_h\left(\begin{smallmatrix}
n& j\\
m& i
\end{smallmatrix}\right)$ being often written as $\lambda\left(\begin{smallmatrix}
n& j\\
m& i
\end{smallmatrix}\right)$ for simplicity.

 For two excitons $i$ and $j$ having opposite momenta, i.e., $i=(\nu_i,\vQ)$ and $j=(\nu_j,-\vQ)$, Eq.~(\ref{app:def_lambda_h}) reduces to
\be
\sum\langle \nu_m|\vk^\prime\rangle\langle \nu_n|\vp^\prime\rangle\langle\vp|\nu_j\rangle\langle\vk|\nu_i\rangle,\label{app:def_lambda_h2}
\ee
provided that the $(\vk^\prime,\vp^\prime,\vk,\vp)$ momenta are such that $-\vp^\prime-\alpha_h\vQ^\prime=-\vk+\alpha_h\vQ$, $-\vk'+\alpha_h\vQ'=-\vp-\alpha_h\vQ$,  $\vk^\prime+\alpha_e\vQ^\prime=\vk+\alpha_e\vQ$, and $\vp^\prime-\alpha_e\vQ^\prime=\vp-\alpha_e\vQ$, as read from Fig.~\ref{fig:lambda_h}. When inserted into Eq.~(\ref{app:def_lambda_h}), we get
\bea
\lefteqn{\lambda_h\left(\begin{smallmatrix}
(\nu_n,-\vQ^\prime)& (\nu_j,-\vQ)\\
(\nu_m,\vQ^\prime)& (\nu_i,\vQ)
\end{smallmatrix}\right)}\nn\\
&=&\sum_\vp\langle \nu_m|\vp+\frac{\vP_-}{2}\rangle\langle \nu_n|\vp-\frac{\vP_-}{2}\rangle\nn\\
&&\times\langle \vp+\frac{\vP_+}{2}|\nu_i\rangle\langle \vp-\frac{\vP_+}{2}|\nu_j\rangle,\label{app:def_lambda}
\eea
where the momenta $\vP_\pm$ are defined by $\vP_\pm=\alpha_h(\vQ+\vQ^\prime)\pm\alpha_e(\vQ^\prime-\vQ)$.

\subsection{Direct Coulomb scattering}
We now turn to the part of the direct Coulomb scattering resulting from hole-hole interaction. As seen from the second diagram of Fig.~\ref{fig:Xi}, it reads as
\bea
\xi_{h_1 h_2}^{\rm dir}\big(\begin{smallmatrix}
n& j\\
m& i
\end{smallmatrix}\big)&=&\sum_{\vq,\vk_e,\vk_h,\vp_e,\vp_h}V_\vq\langle m|\vk_e,\vk_h-\vq\rangle\label{app:def_dirCoul_hh}\\
&&\times\langle n|\vp_e,\vp_h+\vq\rangle\langle\vp_h,\vp_e|j\rangle\langle\vk_h,\vk_e|i\rangle.\nn
\eea
In this direct process, the excitons keep their electron and hole components. As seen from Fig.~\ref{fig:Xi_part}, for two excitons $i=(\nu_i,\vQ)$ and $j=(\nu_j,-\vQ)$, Eq.~(\ref{app:def_dirCoul_hh}) reduces to
\be
\sum V_\vq\langle \nu_m|\vk^\prime\rangle\langle \nu_n|\vp^\prime\rangle\langle\vp|\nu_j\rangle\langle\vk|\nu_i\rangle,\label{app:def_dirCoul_hh2}
\ee
provided that the $(\vk^\prime,\vp^\prime,\vk,\vp)$ momenta are such that $-\vk^\prime+\alpha_h\vQ^\prime=-\vk+\alpha_h\vQ-\vq$, $-\vp^\prime-\alpha_h\vQ^\prime=-\vp-\alpha_h\vQ+\vq$,  $\vk^\prime+\alpha_e\vQ^\prime=\vk+\alpha_e\vQ$, and $\vp^\prime-\alpha_e\vQ^\prime=\vp-\alpha_e\vQ$, as read from the figure. When inserted into Eq.~(\ref{app:def_dirCoul_hh}), this gives
\bea
\lefteqn{\xi_{h_1h_2}^{\rm dir}\Big(\begin{smallmatrix}
(\nu_n,-\vQ^\prime)& (\nu_j,-\vQ)\\
(\nu_m,\vQ^\prime)& (\nu_i,\vQ)
\end{smallmatrix}\Big)}\label{app:def_dirXi_hh}\\
&&=V_{\vP_0}\sum_{\vk\vp}\lan \nu_m|\vk-\alpha_e \vP_0\ran\lan\nu_n|\vp+\alpha_e \vP_0\ran\lan \vk|\nu_i\ran\lan \vp|\nu_j\ran,\nn
\eea
where $\vP_0=\vQ^\prime-\vQ$ is the momentum transfer. Using the same procedure for $\xi_{e_1e_2}^{\rm dir}$, $\xi_{h_1e_2}^{\rm dir}$, and $\xi_{e_1h_2}^{\rm dir}$, we end with a direct Coulomb scattering between two excitons given by
\bea
\lefteqn{\xi^{\rm dir}\left(\begin{smallmatrix}
(\nu_n,-\vQ^\prime)& (\nu_j,-\vQ)\\
(\nu_m,\vQ^\prime)& (\nu_i,\vQ)
\end{smallmatrix}\right)}\label{app:def_dirXi}\\
&=&V_{\vP_0}\sum_{\vk\vp}\Big[\lan \nu_m|\vk+\alpha_h \vP_0\ran\lan\nu_n|\vp-\alpha_h \vP_0\ran\nn\\
&&+\lan \nu_m|\vk-\alpha_e \vP_0\ran\lan\nu_n|\vp+\alpha_e \vP_0\ran\nn\\
&&-\lan \nu_m|\vk+\alpha_h \vP_0\ran\lan\nu_n|\vp+\alpha_e \vP_0\ran\nn\\
&&-\lan \nu_m|\vk-\alpha_e \vP_0\ran\lan\nu_n|\vp-\alpha_h \vP_0\ran\Big]\nn\\
&&\times\lan \vk|\nu_i\ran\lan \vp|\nu_j\ran\nn.
\eea
By noting that
\bea
&&\sum_\vk \big[\lan \nu^\prime|\vk+\alpha_h \vq\ran-\lan \nu^\prime|\vk-\alpha_e \vq\ran\big]\lan \vk|\nu\ran\nn\\
&&=\lan \nu^\prime|e^{i\alpha_h \vq\cdot\vr}-e^{-i\alpha_e \vq\cdot\vr}|\nu\ran=\mathcal{T}_{\nu^\prime\nu}(\vq),
\eea
this direct Coulomb scattering splits as
\be
\xi^{\rm dir}\left(\begin{smallmatrix}
(\nu_n,-\vQ^\prime)& (\nu_j,-\vQ)\\
(\nu_m,\vQ^\prime)& (\nu_i,\vQ)
\end{smallmatrix}\right)=V_{\vP_0}\mathcal{T}_{\nu_m\nu_i}(\vP_0)\mathcal{T}_{\nu_n\nu_j}(-\vP_0).\label{app:dirInt1}
\ee
For $\nu$ and $\nu^\prime$ restricted to the exciton ground state $\nu_0$, we find
\be
\mathcal{T}_{\nu_0\nu_0}(\vq)=g\left(\frac{\alpha_h a_X q}{2}\right)-g\left(\frac{\alpha_e a_X q}{2}\right),
\ee
where, for 2D and 3D systems, $g_{2D}(p)=(1+p^2/4)^{-3/2}$ and $g_{3D}(p)=(1+p^2)^{-2}$. We see that $\mathcal{T}_{\nu_0\nu_0}(\vq)$ depends on the magnitude of the momentum transfer $q=|\vq|$ only and that $\mathcal{T}_{\nu_0\nu_0}(\vq=0)=0$. We also note that $\mathcal{T}_{\nu_0\nu_0}(\vq)=0$ for $\alpha_e=\alpha_h$, i.e., equal electron and hole masses.\

\subsection{Exchange-Coulomb scatterings}

Two excitons can also have exchange-Coulomb scatterings. These are defined as
 \bea
 \xi^{\rm in}\big(\begin{smallmatrix}
n& j\\ m& i\end{smallmatrix}\big)&=&\sum_{rs}\lambda_h\big(\begin{smallmatrix}
n& s\\ m& r\end{smallmatrix}\big)\xi^{\rm dir}\big(\begin{smallmatrix}
s& j\\ r& i\end{smallmatrix}\big),\\
\xi^{\rm out}\big(\begin{smallmatrix}
n& j\\ m& i\end{smallmatrix}\big)&=&\sum_{rs}\xi^{\rm dir}\big(\begin{smallmatrix}
n& s\\ m& r\end{smallmatrix}\big)\lambda_h\big(\begin{smallmatrix}
s& j\\ r& i\end{smallmatrix}\big).
\eea
depending on if the exchange takes place after or before Coulomb interaction.

The part of the ``in" exchange-Coulomb scattering due to hole-hole interaction, as shown in the first diagram of Fig.~\ref{fig:Xi_in}, is given by
\bea
\xi_{h_1 h_2}^{\rm in}\big(\begin{smallmatrix}
n& j\\
m& i
\end{smallmatrix}\big)&=&\sum_{\vq,\vk_e,\vk_h,\vp_e,\vp_h}V_\vq\langle m|\vk_e,\vp_h+\vq\rangle\label{app:def_inexCoul_hh}\\
&&\times\langle n|\vp_e,\vk_h-\vq\rangle\langle\vp_h,\vp_e|j\rangle\langle\vk_h,\vk_e|i\rangle.\nn
\eea
As seen from Fig.~\ref{fig:Xi_in_hh}, for $i=(\nu_i,\vQ)$ and $j=(\nu_j,-\vQ)$, Eq.~(\ref{app:def_inexCoul_hh}) reduces to the same equation as Eq.~(\ref{app:def_dirCoul_hh2}), the $(\vk^\prime,\vp^\prime,\vk,\vp)$ momenta being now such that $\vk^\prime+\alpha_h\vQ^\prime=\vp-\alpha_h\vQ+\vq$, $\vp^\prime-\alpha_h\vQ^\prime=\vk+\alpha_h\vQ-\vq$,  $-\vk^\prime+\alpha_e\vQ^\prime=-\vp-\alpha_e\vQ$, and $-\vp^\prime-\alpha_e\vQ^\prime=-\vk+\alpha_e\vQ$. This leads to
\bea
\lefteqn{\xi^{\rm in}_{ c_1c_2}\left(\begin{smallmatrix}
(\nu_n,-\vQ)& (\nu_j,-\vQ^\prime)\\
(\nu_m,\vQ)& (\nu_i,\vQ^\prime)
\end{smallmatrix}\right)}\nn\\
&=&\sum_{\vk,\vp\not=0}V_\vp\langle \nu_m|\vk+\frac{\vP_-+\vp}{2}\rangle\langle \nu_n|\vk-\frac{\vP_-+\vp}{2}\rangle \nn\\
&&\times\langle \vk+\frac{\vP_+\mp\vp}{2}|\nu_i\rangle\langle \vk-\frac{\vP_+\mp\vp}{2}|\nu_j\rangle,\label{app:inInt1}
\eea
with the lower sign in front of $\vp$ for $c_1c_2=h_1h_2$ and the upper sign for $c_1c_2=e_1e_2$.\

The same procedure for the electron-hole part of the interaction yields
\bea
\lefteqn{\xi^{\rm in}_{ c_1d_2}\left(\begin{smallmatrix}
(\nu_n,-\vQ^\prime)& (\nu_j,-\vQ)\\
(\nu_m,\vQ^\prime)& (\nu_i,\vQ)
\end{smallmatrix}\right)}\nn\\
&=&-\sum_{\vk,\vp\not=0}V_\vp\langle \nu_m|\vk+\frac{\vP_-+\vp}{2}\rangle\langle \nu_n|\vk-\frac{\vP_-+\vp}{2}\rangle\nn\\
&&\times \langle \vk+\frac{\vP_+\mp\vp}{2}|\nu_i\rangle\langle \vk-\frac{\vP_+\pm\vp}{2}|\nu_j\rangle,\label{app:inInt2}
\eea
with the upper sign for $c_1d_2=e_1h_2$ and the lower sign for $c_1d_2=h_1e_2$.\

Note that we can eliminate the sum over $\vp$ in the electron-hole part of the $\xi^{\rm in}_{ c_1d_2}$ scattering, by setting $\vk^\prime=\vk-\vp/2$ and by using Eq.~(\ref{app:X_schrodeq}). We then find
\bea
\lefteqn{\xi^{\rm in}_{ e_1h_2}\left(\begin{smallmatrix}
(\nu_n,-\vQ^\prime)& (\nu_j,-\vQ)\\
(\nu_m,\vQ^\prime)& (\nu_i,\vQ)
\end{smallmatrix}\right)}\nn\\
&=&\sum_{\vk}\left(\va_{\nu_m}-\frac{(\vk+\vP_-/2)^2}{2\mu_X}\right)\langle \nu_m|\vk+\frac{\vP_-}{2}\rangle\nn\\
&&\times \langle \nu_n|\vk-\frac{\vP_-}{2}\rangle\langle \vk+\frac{\vP_+}{2}|\nu_i\rangle\langle \vk-\frac{\vP_+}{2}|\nu_j\rangle.\label{app:inInt21}
\eea
Similarly, by setting $\vk^\prime=\vk+\vp/2$, we find
\bea
\lefteqn{\xi^{\rm in}_{ h_1e_2}\left(\begin{smallmatrix}
(\nu_n,-\vQ^\prime)& (\nu_j,-\vQ)\\
(\nu_m,\vQ^\prime)& (\nu_i,\vQ)
\end{smallmatrix}\right)}\nn\\
&=&\sum_{\vk}\left(\va_{\nu_n}-\frac{(\vk-\vP_-/2)^2}{2\mu_X}\right)\langle \nu_m|\vk+\frac{\vP_-}{2}\rangle\nn\\
&&\times \langle \nu_n|\vk-\frac{\vP_-}{2}\rangle\langle \vk+\frac{\vP_+}{2}|\nu_i\rangle\langle \vk-\frac{\vP_+}{2}|\nu_j\rangle.\label{app:inInt22}
\eea

\end{document}